\documentclass[aps,amsfonts,twocolumn,floats,showpacs,amsmath,amssymb,floatfix,nofootinbib,balancelastpage]{revtex4}

\usepackage{hyperref}
\input epsf
\usepackage{graphics}
\usepackage{amsmath}
\usepackage{color}
\usepackage{dcolumn}
\usepackage{graphicx}

\newcommand{\beqa}{\begin{eqnarray}}
\newcommand{\eeqa}{\end{eqnarray}}

\newcommand{\bn}{\hat{\bf n}}

\newcommand{\beq}{\begin{equation}}
\newcommand{\eeq}{\end{equation}}
\newcommand{\bfl}{{\mathbf{l}}}

\newcommand{\bflp}{{\mathbf{l^{\prime}}}}

\newcommand{\intl}[1]{\int {d^2 l_{#1} \over (2\pi)^2}}

\newcommand{\vsp}{\vphantom{\Big[}\\}

\newcommand{\Tr}{{\rm Tr}}

\newcommand{\da}{d_A}

\newcommand{\bfd}{{\mathbf{d}}}

\newlength{\tskip}
\newlength{\colwidth}

\newcommand{\rad}{r}    % comoving radial distance

\newcommand{\len}{\phi^{len}}

\newcommand{\esttau}{{{\hat\tau}}({\mathbf{\hat{n}}})}
\newcommand{\estphi}{{{\hat\phi}}({\mathbf{\hat{n}}})}

\def\be{\begin{equation}}
\def\ee{\end{equation}}
\def\ba{\begin{eqnarray}}
\def\ea{\end{eqnarray}}
\def\nn{\nonumber}

\def\edth{\;\raise1.0pt\hbox{$'$}\hskip-6pt\partial}
\def\baredth{\;\overline{\raise1.0pt\hbox{$'$}\hskip-6pt
\partial}}

\newcommand{\HII}{H\,{\scriptsize II}}
\newcommand{\HI}{H\,{\scriptsize I}}

\begin{document}

\title{An Improved Forecast of Patchy Reionization Reconstruction with CMB}
%Decoupling Patchy Reionization with Lensing from the Cosmic Microwave Background
\author{Meng Su$^1$}\email{mengsu@cfa.harvard.edu}
\author{Amit P.S. Yadav$^{2}$}
\author{Matthew McQuinn$^{3}$}
\author{Jaiyul Yoo$^{4,5}$}
\author{Matias Zaldarriaga$^{2}$}

\affiliation{$^1$ Institute for Theory and Computation, Harvard-Smithsonian Center for Astrophysics, 60 Garden Street, MS-10, Cambridge, MA 02138}
\affiliation{$^2$ School of Natural Sciences, Institute for Advanced Study, Einstein Drive, Princeton, NJ 08540}
\affiliation{$^3$ Department of Astronomy, University of California, Berkeley, CA 94720}
\affiliation{$^4$ Institute for Theoretical Physics, University of Zurich, Winterthurerstrasse 190, CH-8057 Zurich, Switzerland}
\affiliation{$^5$ Lawrence Berkeley National Laboratory, University of California, Berkeley, CA 94720}
%\affiliation{$^6$ mengsu@cfa.harvard.edu}

\begin{abstract}
% Inhomogeneous reionization gives rise to angular fluctuations in the CMB optical depth $\tau(\bn)$ to the last scattering surface, generating secondary anisotropies in CMB temperature and polarization. We first demonstrate that the gravitational lensing of CMB is the dominant source of contamination for reconstructing inhomogeneous reionization signals, even with $98\%$ of its contribution removed by delensing. Improving upon previous work, we then construct unbiased estimators that can simultaneously reconstruct inhomogeneous reionization signals $\tau(\bn)$ and gravitational lensing potential $\Phi(\bfl)$. We apply our new unbiased estimators to future CMB experiments to assess the detectability of inhomogeneous reionization signals. With more physically motivated models for reionization that predict an order of magnitude smaller signals than previous studies, we show that a CMBPol-like experiment could achieve a marginal detection of patchy reionization, $(S/N)2 \sim O(1)$ to $\sim O(10)$.

Inhomogeneous reionization gives rise to angular fluctuations in the Cosmic Microwave Background (CMB) optical depth $\tau(\bn)$ to the last scattering surface, correlating different spherical harmonic modes and imprinting characteristic non-Gaussianity on CMB maps.  Recently the minimum variance quadratic estimator $\hat\tau(\bn)$ has been derived using this mode-coupling signal, and found that the optical depth fluctuations could be detected with $(S/N)^2\sim 100$ in futuristic experiments like CMBPol. We first demonstrate that the non-Gaussian signal from gravitational lensing of CMB is the dominant source of contamination for reconstructing inhomogeneous reionization signals, even with $98\%$ of its contribution removed by delensing.  We then construct unbiased estimators that simultaneously reconstruct inhomogeneous reionization signals $\tau(\bn)$ and gravitational lensing potential $\phi(\bn)$.  We apply our new unbiased estimators to future CMB experiment to assess the detectability of inhomogeneous reionization signals.  With more physically motivated simulations of inhomogenuous reionizations that predict an order of magnitude smaller $C_{l}^{\tau\tau}$ than previous studies, we show that a CMBPol-like experiment could achieve a marginal detection of inhomogeneous reionization, $(S/N)^2 \sim\mathcal{O}(1)$ with this quadratic estimator to $\sim\mathcal{O}(10)$ with the analogous maximum likelihood estimator.
\end{abstract}
\maketitle

\section{Introduction}

Reionization marks the time in which the vast majority of the hydrogen in the Universe was ionized.  When and how this process occurred is at present poorly constrained.  Current data show that it must have ended by $z \approx 6$ because at lower redshifts there was significant transmission in the Ly$\alpha$ forest \citep{Fan2006}.  In addition, the large-scale polarization anisotropies in the cosmic microwave background (CMB) constrain the mean redshift of reionization to have been $z = 10.6\pm 1.2$ \citep{wmap7_cosmology}.

It is believed that the first galaxies in the Universe produced the ionizing photons that ultimately ionized the intergalactic gas (e.g. \cite{BarkanaLoeb2001}).  The morphology of reionization and its duration depended on the nature, abundance, and clustering of the ionizing sources \citep{Furlanetto2004, Mcquinn2007}.   There are several established ideas for how to better constrain the morphology and the redshifts over which it occurred.  These include detecting the reionization-induced suppression and spatial modulation in the statistics of high-redshift Lyman-$\alpha$ emitting galaxies \citep{miralda98,mcquinn07, ouchi10}, studying \HI\ Lyman-$\alpha$ damping wing absorption from the neutral gas during reionization in the afterglow spectra of high-redshift gamma ray bursts \citep{miralda98, totani06, mcquinn08}, and directly observing $21~$cm emission from $z>6$ neutral hydrogen in the intergalactic medium (e.g., \citet{furlanetto06}).  This study concentrates on using a new technique, first proposed in \citet{DvorkinSmith}, that exploits the non-Gaussianities in the CMB sourced by reionization to study this process.
%Unlike other probes, the CMB is the only diagnostic that becomes more sensitive to the signal as the redshifts over which reionization happened increase.

Inhomogeneous reionization produces several secondary anisotropies in the CMB.  First, extra temperature (and, to a lesser extent, polarization) anisotropies are generated from peculiar motion of ionized regions during the entire reionization process ~\cite{SZ1970,SZ1980, Hu2000, McQuinn2005,Santos2003,Iliev2007,Zhang2004}.  These anisotropies are termed the kinetic Sunyaev-Zeldovich effect.  Second, ionized bubbles scatter the local CMB temperature quadrupole, generating fluctuations in the polarization at {\it large scales}~\cite{GruzinovHu,Hu2000}. Finally, the patchy nature of reionization would have resulted in the Thomson scattering optical depth to recombination, $\tau(\bn)$, depending on direction ~\cite{Dore2007,McQuinn2005,Hu2000,Aghanim2008,holder2009}. Such optical depth fluctuations act as a modulation effect on CMB fields by suppressing the primordial anisotropies with a factor of $e^{-\tau(\bn)}$, correlating different spherical harmonics. Information contained in $\tau(\bn)$ fluctuations could potentially probe the duration of hydrogen reionization and the size of the ionized regions. 

It is well known that gravitational lensing also imprints a non-Gaussian signature on the CMB. Minimum variance quadratic estimator has been introduced by using the coupled modes to reconstruct the projected lensing potential~\cite{HuOkamoto,OkamotoHu2003,HirataSeljak2003,LewisChallinor2006}. Recently, \citet{DvorkinSmith} followed similar technique and derived the minimum variance quadratic estimator for $\tau(\bn)$.  Utilizing a toy model for reionization, they estimated that the patchy reionization signal could be detected with $(S/N)^2 \sim 100$ for a CMBPol-like experiment, with beam full-width half-maximum (FWHM) of $\Theta_{\rm FWHM}=4'$, and noise sensitivity $\Delta_T=1 \mu \text{k-arcmin}$. In this paper, we quantify the impact of lensing induced non-Gaussianities on the reconstruction of $\tau(\bn)$. We show that lensing biases the reconstruction of $\tau(\bn)$, and as a solution we construct an unbiased estimator for $\tau(\bn)$ in the presence of lensing.  

The structure of this paper is as follows. Section~\ref{sec:simu} provides simple estimates for the size of $\tau (\bn)$ fluctuations, and it describes the cosmological reionization calculations used here to produce $\tau (\bn)$ maps.  Section~\ref{sec:esti} derives the minimum variance quadratic estimator for $\hat\tau(\bn)$ in the flat sky approximation, and it quantifies the effect on lensing on the estimator.  %This motivates constructing an estimator that is not biased by lensing.
 In section~\ref{sec:nume}, we summarize our results and discuss the implications.  
In Appendix~\ref{sec:simulation}, we describe in more detail our simulations to reconstruct $\tau(\bn)$ in the presence of lensing.

\begin{table}[t]
\caption{Description of various reionization models.  Distances are in comoving units.}
\begin{center}
\begin{tabular}{c|c|c|c|c|c|c}
  & $\zeta_7$ & $d\zeta/dz$ & $l_{\rm mfp}$ (Mpc)& $\tau$& $L_{\rm box}$ (Mpc) & $N_g$ \\
\hline 
A & 10 & 0  & $\infty$ & $0.063$ & 200 & 256 \\
B & 10 & 18 & $\infty$ & $0.112$  & 200 & 256 \\
C & 30 & 0  & $10$    & $0.090$  & 200 & 256 \\
D & 20 & 16 & $10$    & $0.115$  & 200 & 256 \\
\hline
\end{tabular}
\end{center}
\label{taumodel}
\end{table}

\section{Inhomogeneous Reionization}
\label{sec:simu}

Patchy reionization produced a line-of-sight dependent optical depth that can be written as
\begin{equation}
\tau(\hat{\bf n}) = c \int \frac{a \, dz}{H(z)} \sigma_T \, \bar{n}_e(z) \, \left[1 + \delta_b(\hat{\bf n}, z) + \delta_x(\hat{\bf n}, z) \right], 
\end{equation}
where $\sigma_T$ is the Thompson scattering optical depth, $\bar{n}_e(z)$ is the average free-electron proper number density, and $\delta_b$ and $\delta_x$ are the over-densities in baryons and in the ionized fraction, $x_i$.

The angular power spectrum $C_{l}^{\tau \tau}$ in the flat sky approximation can be related to the $3$D ionization and density field using the Limber approximation:
\begin{eqnarray}
C_{l}^{\tau \tau} &=& \int \frac{d \eta}{\eta^2} \, a^{2} \, \sigma_T^2 \, \bar{n}_e(z)^2 \, \bigg[P_{xx}(z, \frac{\bfl}{\eta}) \nonumber \\
&+&  2 \,P_{x \delta}(z, \frac{\bfl}{\eta}) 
 +   P_{\delta \delta}(z, \frac{\bfl}{\eta}) \bigg],
 \label{eqn:ctau}
\end{eqnarray}
where $\eta$ is the conformal distance from the observer, and $P_{XY}$
is the cross power spectrum of the over-density in $X$ with the
overdensity in $Y$.  This power spectrum is weighted heavily to the
highest redshifts where there was reionized gas. The kinetic Sunyaev-Zeldovich effect (kSZ) signal from reionization is predicted to be comparable to the
kSZ signal after reionization.  However, the kSZ weights by an additional $v^2$ factor which goes roughly as the scale factor, result in its kernel
peaking at lower redshifts~\cite{Hu2000}. This results in a large
fraction of the kSZ coming from after reionization, we can safely
neglect the low redshift part, whereas we expect most of $C_{l}^{\tau \tau}$ to originate from during reionization.  It is also clear from equation (\ref{eqn:ctau}) that $C_{l}^{\tau \tau}$ from reionization increases approximately linearly with the duration of reionization for fixed mean redshift of reionization.

It is likely that reionization occurred in a patchy manner, with some
regions being ionized early on in this process and others remaining
neutral until the end, and with little gas at intermediate ionization
states.  This patchiness likely resulted in the ionization
fluctuations dominating over other sources of fluctuation (i.e.,
$P_{xx} \gg P_{\delta \delta}$ on arcmin and larger scales~\cite{Furlanetto2004}).  Even without any knowledge of $P_{xx}$ other than that reionization was patchy, there is an integral
constraint on $C_{l}^{\tau \tau}$ because if the ionization field is
zeros and ones $\int k^2dk/(2\pi^2)  \, P_{xx} = x_i^{-1} -1$, where
$x_i$ is the ionized fraction.  Thus, fixing the reionization history
and in the Limber approximation, $\int l^2 dl \, C_{l}^{\tau \tau}$ is
just a single number independent of morphology.  This constraint shows
that the larger the \HII\ regions during reionization, the larger the
fluctuations in $\tau(\bn)$\footnote{ Since $l^3 C_{l} \sim
  const.$, if the peak in the bubble scale is at smaller $l$
  (i.e. larger bubbles), as $l^2 C_{l} \sim l^{-1}$, larger bubbles
  result in a higher peak (but at lower $l$ i.e. larger fluctuations).}.

To estimate $C_{l}^{\tau \tau}$, we compute Monte-Carlo realizations of reionization in two hundred comoving Mpc data cubes using 
the method developed in~\cite{zahn2007} for assigning the ionization state to boxes with realizations of the linear-theory cosmological density field.  This method is based on the semi-analytic model for reionization in~\cite{Furlanetto2004}.  The distribution of ionized gas found in the \cite{zahn2007} method is in excellent agreement 
with the results of detailed numerical simulations of reionization~\cite{zahn2007, zahn2010}.  Thus, we expect that the $\tau$ field from this simulation will be more realistic than the analytic model used in the original study of \cite{DvorkinSmith}.  Their model assumed a lognormal distribution of bubbles with a distribution that was independent of ionized fraction.  In our calculations, the morphology of the bubbles is complicated and their sizes increase dramatically as $x_i$ increases.

The method in \cite{zahn2007} 
that we employ posits that the number 
of galaxies within a region sets its ionization state.  Namely,
a region is ionized if $1 > \zeta \, f$, where $\zeta$ is a factor that 
encodes the efficiency that galaxies can ionize their surroundings, and $f$ is 
the total fraction that has collapsed into halos with mass $>$ ${\it m}_{\rm min}$, where ${\it m}_{\rm min}$ is the minimum halo mass of the sources during reionization.  A point in space is marked as ionized 
if this criterion is met for any smoothing scale centered around it (where the smoothing is done with a tophat in Fourier space filter).

Calculating $f$ in detail requires high-resolution $N$-body simulations to 
resolve $\sim10^8 \, M_{\odot}$ halos -- the smallest halos that were expected to form multiple generations of stars --, while still capturing scales much larger than the 10 comoving Mpc bubbles. 
Fortunately, extended Press-Schechter theory provides a method to
calculate $f$ in a macroscopic region of size R and overdensity $\delta_R$ in the simulations from just the linear 
density field~\cite{BCEK,LaceyCole}.
Therefore, we can quickly compute the ionization field from the linear density field and rather course resolution 
using just Fast Fourier Transforms.  Our calculations take just minutes on a single CPU for the $256^3$ grids used here. % Furthermore, the ionization field generated using this method is in almost as good agreement with numerical simulations of reionization as methods in which f is computed from N -body simulations~\cite{zahn2007}.

%%%%%%%%%%%%%%%%%%%%%%%%%%%%%%%%%%%%%%%%%%%%%%%%%%%%%%%%%%%%%%%%%%%
%%%%%%%%%%%%%                 FIGURE 3
%%%%%%%%%%%%%%%%%%%%%%%%%%%%%%%%%%%%%%%%%%%%%%%%%%%%%%%%%%%%%%%%%%%
\begin{figure}
\includegraphics[width=85mm,angle=0]{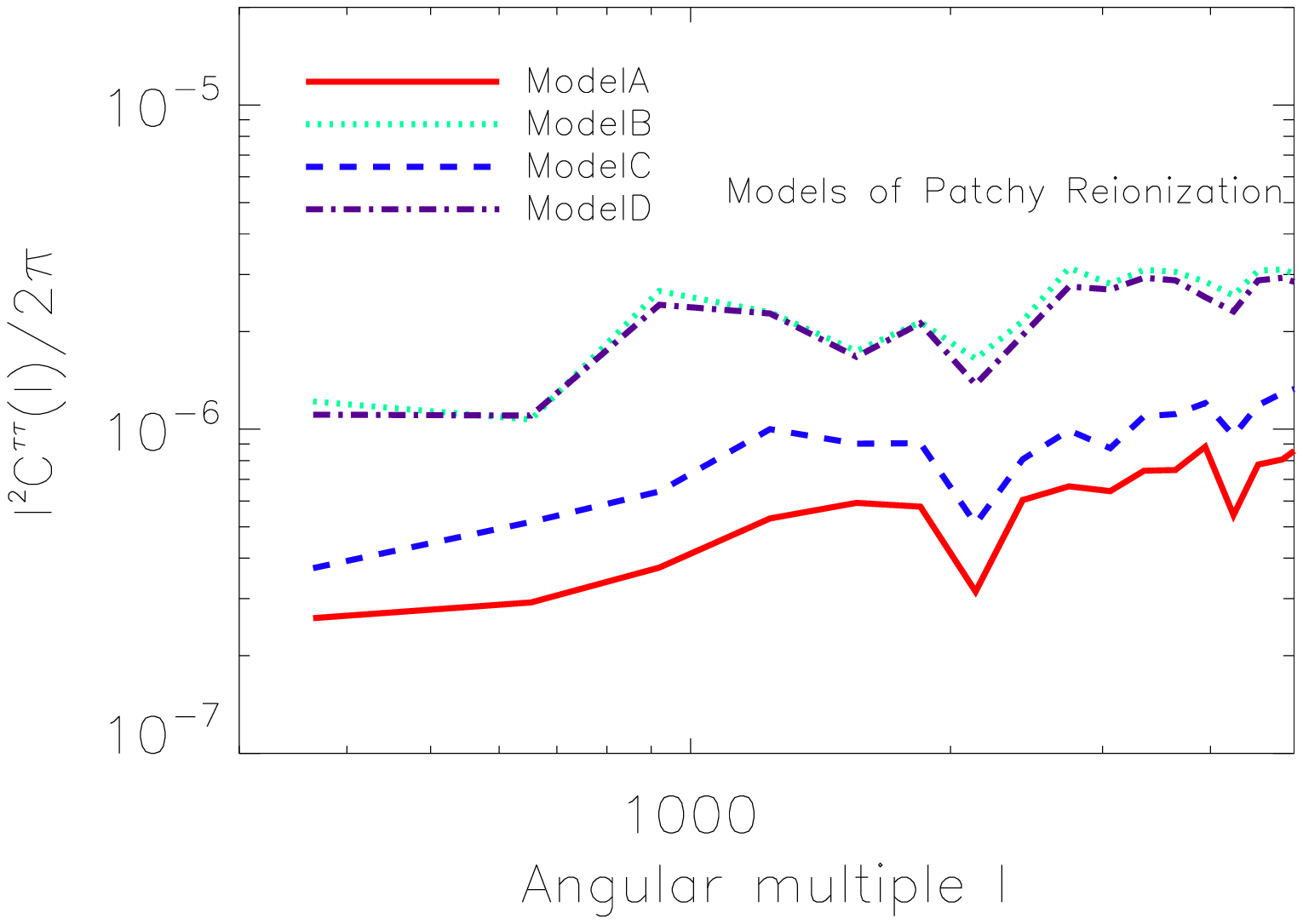}
\includegraphics[width=85mm,angle=0]{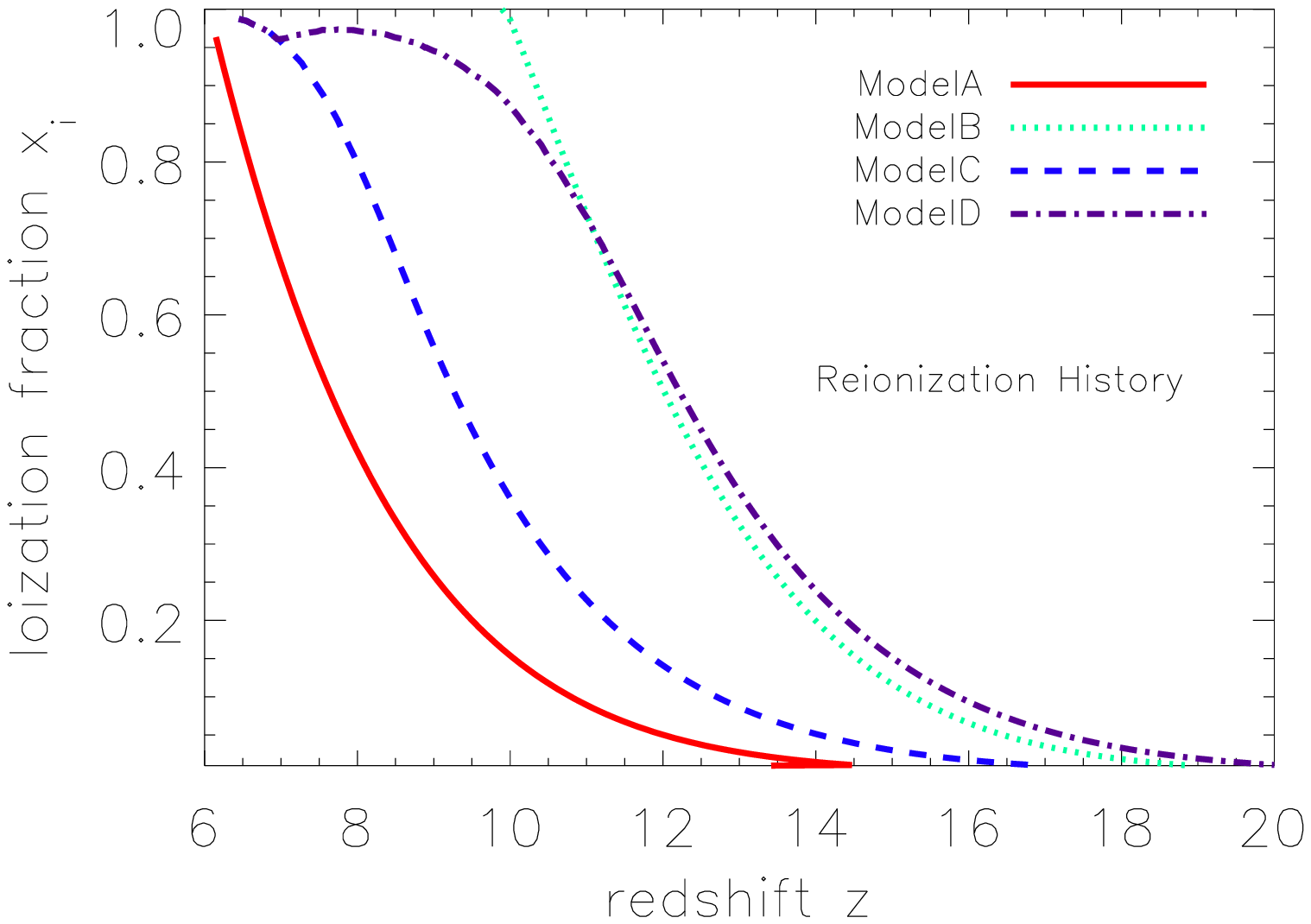}
\caption{ {\it Upper panel}:  Optical depth power spectrum from the different 
reionization models described in Table~\ref{taumodel}.  {\it Lower panel}: Corresponding 
average reionization history of these models. The mean optical depth of
each model is consistent with the WMAP measurement ($\tau=0.088\pm0.015$).}
\label{fig:tausim}
\end{figure}
%%%%%%%%%%%%%%%%%%%%%%%%%%%%%%%%%%%%%%%%%%%%%%%%%%%%%%%%%%%%%%%%%%%

There is significant uncertainty in the properties of the first sources and sinks 
that were responsible for reionization.  All of our models assume that the ionizing luminosity is proportional to the collapsed mass in halos above $10^8~M_\odot$ (approximately the minimum mass threshold where the gas can cool by atomic transitions and form dense structures).  To explore the allowed parameter space 
of this process, we model reionization with 3 parameters: the ionizing 
efficiency of a halo $\zeta_7$ at $z = 7$, its derivative with redshift $d\zeta/dz$
(assumed to be independent of $z$), and
the mean free path of ionizing photons to be absorbed by an over-dense sink $l_{\rm mfp}$ within an ionized region.  
The first two parameters primarily affect the duration of reionization while
 the later parameter primarily affects its
morphology~\cite{Furlanetto2005,Mcquinn2007}. 

In particular, bubbles that are larger than mean free path of ionizing photons have most of the photons 
produced within them absorbed by dense systems within the bubble
rather than by diffuse gas at the bubble edge, preventing further growth~\cite{Furlanetto2005,Mcquinn2007}. Thus, the parameter $l_{\rm mfp}$ is implemented by setting 
the maximum smoothing scale used to be $l_{\rm mfp}$.
We generate Monte-Carlo maps for 4 different reionization models, which are described in Table~\ref{taumodel}.  All of our models fall within $2\sigma$ of the best fit WMAP $\tau$ 
measurement of $\tau = 0.088 \pm 0.015$~\cite{wmap7_cosmology}.  

In Fig.~\ref{fig:tausim}, the top panel shows 
the optical depth fluctuation power spectrum of the different reionization models described in Table~\ref{taumodel}.  The corresponding reionization history of the four models are also shown in the bottom panel of Fig.~\ref{fig:tausim}.  Surprisingly, the spectrum of all these models is not significantly different:  All the models scales as $l^2 C^{\tau \tau}_{l}/2\pi\approx$ constant for $200\lesssim l\lesssim 10000$.  However, the amplitude varies between $\sim 10^{-6}- 10^{-7}$, owing to the different reionization histories.  An amplitude of $10^{-6}$ is still an order of magnitude smaller that the signal considered in the previous work of~\cite{DvorkinSmith}.  It is possible that reionization is more extended than in our models.  We note that the amplitude of $C^{\tau \tau}_{l}$ is proportional to the duration of reionization.\footnote{Recently it was shown that the velocity difference between the baryons and dark matter that is imparted up until recombination and decays away thereafter, can suppress the formation and baryonic accretion of the $\lesssim 10^6~M_\odot$ halos that harbor the first stars \cite{tseliakhovich10, dalal10}.  The standard paradigm is that these halos did not contribute significantly to reionization \citep{furlanetto06}, but they may have ionized the intergalactic medium fractionally.  Different regions in the Universe have different velocity offsets, with the coherence length of this difference being hundreds of Mpc.  Even if these first stars just fractionally ionized the Universe, this large-scale modulation of the velocity difference could lead to larger fluctuations in $\tau(\hat n)$ (and peaking at $l \sim 100$) than in the models we have considered.  Thus, we point out that there remains the possibility of generating a larger signal than in the models considered here.}  
%{\bf [It would be nice to have a quick estimate for how large this could be using the integral constrain mentioned earlier]}

%%%%%%%%%%%%%%%%%%%%%%%%%%%%%%%%%%%%%%%%%%%%%%%%%%%%%%%%%%%%%%%%%%%

\begin{table*}[t]
\caption{Minimum variance filters for optical depth estimator $\esttau$ and lensing potential estimator $\hat\phi(\bn)$ }
\begin{center}
\begin{tabular}{c|c|c}
\hline \hline
$X X'$                & $f^{\tau}_{XX'}({\bfl_1,\bfl_2})$   & $f^{lens}_{XX'}({\bfl_1,\bfl_2})$\vsp \hline
$T T$      & $C_{l_1}^{T T} + C_{l_2}^{T T}$  & $C_{l_1}^{T T}(\bfl \cdot \bfl_1) + C_{l_2}^{T T}(\bfl \cdot \bfl_2) $\vsp 
$T E$      & $\tilde C_{l_1}^{T E}\cos 2(\varphi_{\bfl_1}-\varphi_{\bfl_2}) + C_{l_2}^{T E}$ & $\tilde C_{l_1}^{T E}(\bfl \cdot \bfl_1)\cos 2(\varphi_{\bfl_1}-\varphi_{\bfl_2}) + C_{l_2}^{T E}(\bfl \cdot \bfl_2)$\vsp
$T B$      & $\tilde C_{l_1}^{T E}\sin 2
                        (\varphi_{\bfl_1}-\varphi_{\bfl_2})$ & $\tilde C_{l_1}^{T E}(\bfl \cdot \bfl_1)\sin 2
                        (\varphi_{\bfl_1}-\varphi_{\bfl_2})
                  $\vsp
$E E$          
                & $[\tilde C_{l_1}^{E E}
                  +\tilde C_{l_2}^{E E}]\cos 2
                        (\varphi_{\bfl_1}-\varphi_{\bfl_2}) $ & $[\tilde C_{l_1}^{E E}(\bfl \cdot \bfl_1)
                  +\tilde C_{l_2}^{E E}(\bfl \cdot \bfl_2)
                        ]\cos 2
                        (\varphi_{\bfl_1}-\varphi_{\bfl_2})$
                         \vsp
$E B$          
                & $[\tilde C_{l_1}^{E E}
                 -\tilde C_{l_2}^{B B}]
                \sin 2(\varphi_{\bfl_1}-\varphi_{\bfl_2})
                $ & $[\tilde C_{l_1}^{E E}(\bfl \cdot \bfl_1)
                 -\tilde C_{l_2}^{B B}(\bfl \cdot \bfl_2)]
                \sin 2(\varphi_{\bfl_1}-\varphi_{\bfl_2})$
                \vsp
$B B$           
                & $[\tilde C_{l_1}^{B B}
                 + \tilde C_{l_2}^{B B}]
                        \cos 2(\varphi_{\bfl_1}-\varphi_{\bfl_2})$
& $[\tilde C_{l_1}^{B B}(\bfl \cdot \bfl_1)
                 + \tilde C_{l_2}^{B B}(\bfl \cdot \bfl_2)]
                        \cos 2(\varphi_{\bfl_1}-\varphi_{\bfl_2})$
                 \vsp
\hline\hline

\end{tabular}
\end{center}
\label{taufilter}
\end{table*}

                                 %%%%%%%%%%%%%%%%%%%%%%%%%%%%%%%
                                 %%%%      SECTION  BEGIN    %%%
                                 %%%%%%%%%%%%%%%%%%%%%%%%%%%%%%%

\section{Standard Quadratic Estimator of  Patchy reionization from the CMB}
\label{sec:esti}

%In section~\ref{sec:tau}, we introduce the simple quadratic estimator for the optical depth field $\tau(\bn)$. In section~\ref{sec:lensing}, we show how lensing would induce significant spurious signal with applying the $\esttau$ estimator to lensed CMB fields. In section~\ref{sec:unbiase}, we describe the unbiased estimator for $\tau(\bn)$ for which lensing would only enhance the estimator noise without producing fake signal.

%\subsection{The Quadratic Estimator of $\tau(\bn)$: Dvorkin \& Smith (2009)}
\label{sec:tau}
The observed CMB temperature and polarization Stokes parameters in the presence of inhomogeneous screening caused by patchy ionizated regions are 
\ba T(\bn)&=&e^{-\delta\tau(\bn)}\tilde
T(\bn) \,, \nn \\ (Q\pm iU)(\bn) &=&e^{-\delta\tau(\bn)} (\tilde Q \pm
i\tilde U)(\bn) \,, 
\label{modulation}
\ea 
where tildes signify the CMB Stokes parameters for a uniform reionization history with constant factor $e^{-\tau}$ spatially modulating the observed CMB fields. We take $\tau$ as the mean of optical depth and $\delta\tau(\bn)$ as the line of sight dependent optical depth fluctuation field. We work in the flat-sky limit where scalar fields such as the CMB temperature $T$ and a complex field $({\cal S}_1 \pm i {\cal S}_2)(\bn)$ of spin $\pm s$ can be expanded in the Fourier basis as
% such as CMB polarization Stokes parameter $(Q \pm i U)(\bn)$ 
\begin{eqnarray}
T(\bfl) &=& \int d \bn \, T(\bn) e^{-i \bfl \cdot \bn}\,, \\
\left[{\cal S}_1 \pm i {\cal S}_2 \right] (\bfl) &=& (\pm 1)^s \int  d \bn\, [{\cal S}_1(\bn)\pm i {\cal S}_2(\bn)] e^{\mp si\varphi_{\bf l}} e^{-i \bfl \cdot \bn}, \nonumber
\label{EBFields}
\end{eqnarray}
where $\varphi_{\bfl}=\cos^{-1}({\bn} \cdot \hat \bfl)$. The complex field $(Q\pm iU)(\bn)$ is a spin $\pm 2$ field, whose Fourier harmonics are referred as $(E\pm iB)(\bfl)$.  
 
Since the differential optical depth fluctuation is already constrained to be small, we work out the effects to first order in $\delta\tau(\bn)$. We use $\tau(\bn)$ rather than $\delta\tau(\bn)$ to specify the fluctuations of optical depth for short. It is simple to show that patchy reionization induces modulations in observed CMB fields that is proportional to $\tau(\bn)$ to the first order. The effect of such a modulation is to correlate different CMB modes in Fourier space. The correlations can be compactly written as
\begin{equation}
\langle X(\bfl_1) X'(\bfl_2) \rangle_{\rm CMB} =  f^{\tau}_{XX'}(\bfl_1,\bfl_2) \, \tau(\bfl)\,,
\label{BasicDifference}
\end{equation}
where $X,X'={T,E,B}$, $\bfl = \bfl_1 + \bfl_2$, $f^{\tau}_{XX'}(\bfl,\bflp)$ is given in Table~\ref{taufilter}, and $\langle ... \rangle_{\rm CMB}$ signifies an ensemble average over CMB realizations with fixed $\tau(\bn)$ field.

The presence of $\tau(\bn)$ field breaks the rotational symmetry of the CMB field, correlating different modes which are not correlated assuming a Gaussian CMB field. Following~\cite{HuOkamoto}, we construct a minimum variance quadratic estimator $\hat{\tau}_{XX'}(\bn)$ for $\tau(\bn)$ field, or $\hat{\tau}_{XX'}(\bfl)$ for $\tau(\bfl)$ in Fourier space. 
\begin{eqnarray}
\hat \tau_{XX'}({\bfl})&=& N_{XX'}(l) \intl{1}
\big[X(\bfl_1) X'(\bfl_2)\big]F^{\tau}_{XX'}(\bfl_1,\bfl_2)\,, \nonumber
\label{eqn:estimator}
\end{eqnarray}
where $\bfl=\bfl_2 + \bfl_1$ and
\begin{eqnarray}
N_{XX'}(l) = \Bigg[ \intl{1} f^{\tau}_{XX'}(\bfl_1,\bfl_2)
F^{\tau}_{XX'}(\bfl_1,\bfl_2) \Bigg]^{-1}.
\label{eq:noise}
\end{eqnarray}
We derive the optimal $F_{XX'}$ by minimizing 
the variance of $\langle \hat \tau_{XX^\prime}(\bfl) \hat\tau_{XX^{\prime}}(\bfl^{\prime})\rangle$.  For $XX'$ = $EE$, $BB$, and $TT$, 
\begin{equation}
F^{\tau}_{XX}(\bfl_1,\bfl_2) = {f^\tau_{XX}(\bfl_1,\bfl_2) \over
	 2 C_{l_1}^{XX,t} C_{l_2}^{XX,t}}.
\label{filter1}
\end{equation}
For $XX'$ = $T B$ and $E B$,
\begin{equation}
F^{\tau}_{XX'}(\bfl_1,\bfl_2) = {f^\tau_{XX'}(\bfl_1,\bfl_2) \over 
	C_{l_1}^{XX,t} C_{l_2}^{X'X',t}},
\label{filter2}
\end{equation}
%\begin{equation}
%F_{XX'}(\bfl_1,\bfl_2) = \frac{f^\tau_{XX'}(\bfl_1,\bfl_2)}{C_{l_1}^{XX,t} C_{l_2}^{X'X',t}}\,,
%\end{equation}
where %$C_{l_2}^{XX,t}$ and $C_{l_2}^{X'X',t}$ are the observed power spectra including the effects of both the signal and the noise,
\begin{eqnarray}
C^{XX',t}_l=\tilde C^{XX'}_l + C^{XX',n}_l\,,
\end{eqnarray}
and $ C^{XX',n}_l$ is the noise power spectrum. We assume the detector noise is Gaussian and isotropic, to be known a priori.  Furthermore, we assume a symmetric Gaussian instrumental beam so that the noise power spectrum is 
\begin{equation}
C_l^{XX,n} = \Delta^2_X e^{l^2\Theta^2_{\rm fwhm}/(8 \, \ln 2) },
\end{equation} 
where $\Delta_X$ is the instrument noise for temperature $(X=T)$ or polarization $(X=E,B)$, and $\Theta_{\rm fwhm}$ is the full-width half-maximum (FWHM) of the Gaussian beam.  We assume a fully polarized detector, for which $\Delta_{E,B}=\sqrt{2}\,\Delta_T$.

The variance of the minimum variance quadratic estimator is
\begin{equation}
\langle \tilde \tau_{XX^\prime}(\bfl_1) \tilde \tau_{XX^{\prime}}(\bfl_2)\rangle=(2\pi)^2 \delta(\bfl_1 + \bfl_2) \{ C^{\tau \tau}_l +N_{XX^\prime}(l)\},
\label{eqn:variance}
\end{equation}
where $N_{XX^\prime}(l)$ gives the dominant contribution to the variance for the $EB$ and $TB$ estimators.

%%%%%%%%%%%%%%%%%%%%%%%%%%%%%%%%%%%%%%%%%%%%%%%%%%%%%%%%%%%%%%%%%%%
%%%%%%%%%%%%%                 FIGURE 2
%%%%%%%%%%%%%%%%%%%%%%%%%%%%%%%%%%%%%%%%%%%%%%%%%%%%%%%%%%%%%%%%%%%
\begin{figure*}[t]
\includegraphics[width=160mm,angle=0]{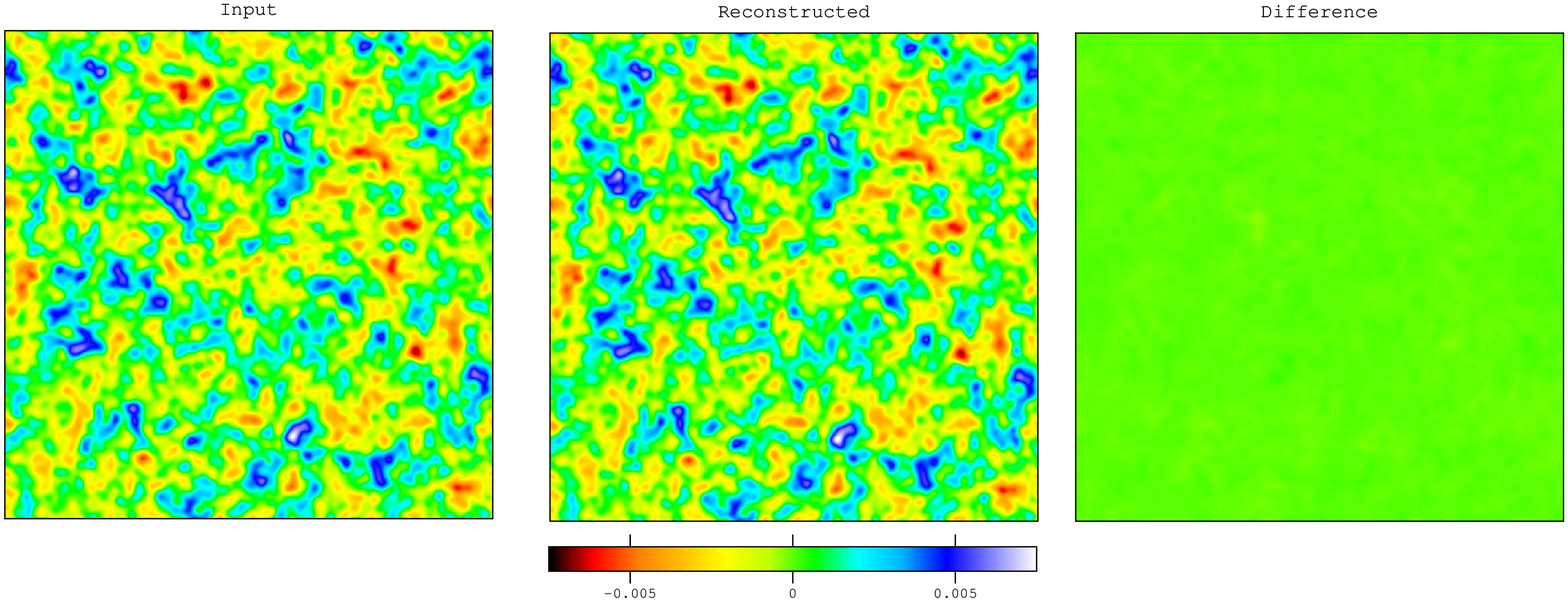}
\includegraphics[width=160mm,angle=0]{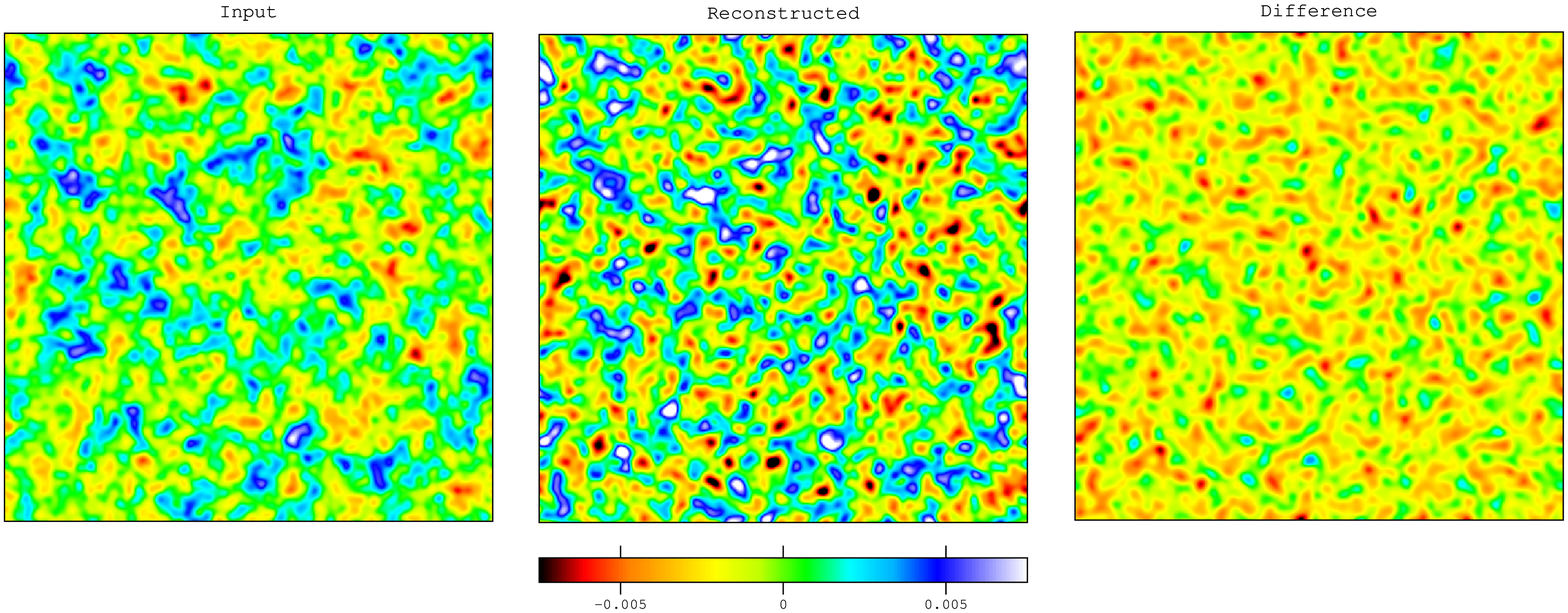}
\caption{Impact of lensing on the reconstruction of the optical depth fluctuation field $\tau(\bn)$. The {\it upper panels} assume all the mode-coupling of CMB maps are generated by modulations of patchy reionization only. The upper left panel shows the input $\tau(\bn)$ map used to modulate the CMB fields, the upper middle panel shows the reconstructed $\tau(\bn)$ map from the CMB fields by applying the quadratic $EB$ estimator, and the upper right panel shows the difference between the input/reconstructed $\tau(\bn)$ maps. The {\it lower panels} show the same quantities as the {\it upper panels} but with the lensing effect on the CMB maps included. As is clear from the lower middle and lower right panels, additional mode-coupling of CMB fields due to lensing contaminates the $\tau(\bn)$ reconstruction. In this plot, we consider the lensing signal to be only $2\%$ of the fiducial value (i.e. the deflection angle power spectrum used here is $C^{dd}_l/50$) to approximate the residual lensing signal after applying delensing procedure on the observed CMB maps. The reconstructed maps were averaged over 1000 CMB realizations for a fixed optical depth fluctuation $\tau(\bn)$ field. Each map is 6$\times$6 square degrees.}
\label{fig2}
\end{figure*}
%%%%%%%%%%%%%%%%%%%%%%%%%%%%%%%%%%%%%%%%%%%%%%%%%%%%%%%%%%%%%%%%%%%

%%%%%%%%%%%%%%%%%%%%%%%%%%%%%%%%%%%%%%%%%%%%%%%%%%%%%%%%%%%%%%%%%%%
%%%%%%%%%%%%%                 FIGURE 3
%%%%%%%%%%%%%%%%%%%%%%%%%%%%%%%%%%%%%%%%%%%%%%%%%%%%%%%%%%%%%%%%%%%
\begin{figure}
%\begin{center}
\includegraphics[width=60mm,angle=-90]{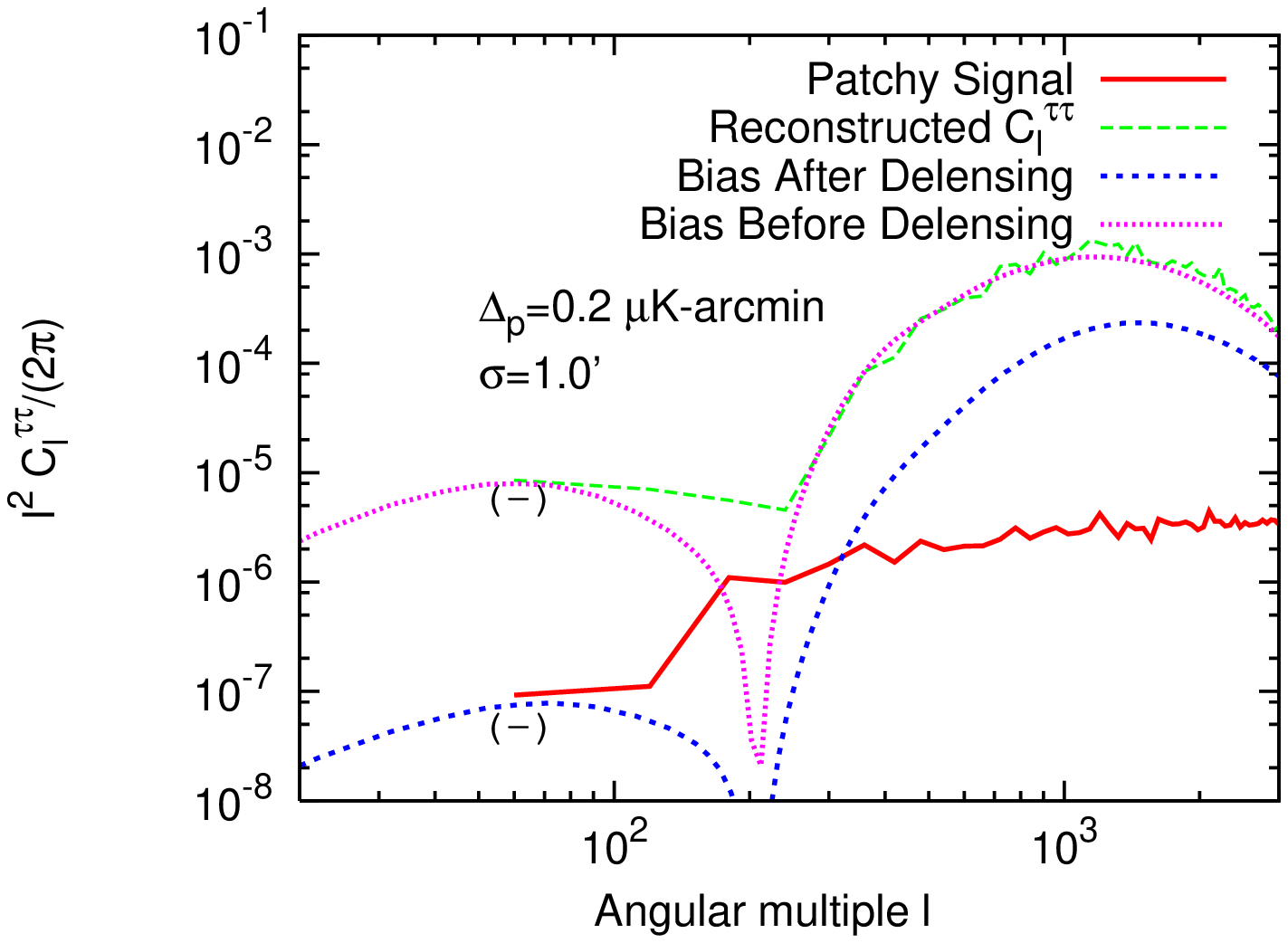}
\caption{Comparison of the biased reconstruction of patchy reionization power spectrum $\hat C_{l}^{\tau\tau}$ with the input fiducial patchy reionization induced power spectrum $C_{l}^{\tau\tau}$. The solid red line shows the $C_{l}^{\tau\tau}$ of the reionization scenario we chose. The dot magenta curve shows the biased estimate for $C_{l}^{\tau\tau}$, which was calculated using Eq.~(\ref{eq:bias}). The lensing bias is negative at low $l$ but positive at high $l$. The minimum of the bias at $l\sim200$ is where the sign changes. The green dashed curve is the same quantity extracted from simulations after averaging over $1000$ realizations.  Our analytic expression matches well with the simulation. The blue short-dashed curve shows the reconstructed $C_l^{\tau\tau}$ after delensing is applied.  The delensing was performed using the quadratic minimum variance estimator for projected lensing potential. 
%{\it Right panel}:  Signal-to-noise as function of maximum multipole. The red solid line shows the $(S/N)^2$ to detect the spurious signal from lensing, which is much larger than the $(S/N)^2$ for this estimator to detect $C_{l}^{\tau\tau}$.  The green dashed line shows the biased signal after delensing. The blue short dashed shows the real patchy signal is order $10$ for ${\rm l} \sim 1000$. 
%We can see that the biased signal has a $(S/N)^2$ is $\sim\mathcal{O}(100)$ times larger than the real signal even after delensing with l up to 1000. 
}
\label{reconstructedPower} 
%\end{center}
\end{figure}

%%%%%%%%%%%%%%%%%%%%%%%%%%%%%%%%%%%%%%%%%%%%%%%%%%%%%%%%%%%%%%%%%%%
%%%%%%%%%%%%%%%%%%%%%%%%%%%%%%%%%%%%%%%%%%%%%%%%%%%%%%%%%%%%%%%%%%%

\section{Lensing Contamination in $\tau(\bn)$  Reconstruction}
\label{sec:lensing}
 
The optical depth estimators described in the previous section neglect the effect of CMB lensing. In reality, both the CMB temperature and polarization fields are gravitationally lensed by inhomogeneities in the matter distribution between the last scattering surface and $z=0$. In this section, we show that lensing significantly bias the $\tau(\bn)$ reconstruction. 

Both the $\tau(\bn)$ field and the projected lensing potential $\len(\bn)$ can generate non-Gaussianity by mixing modes and break the rotational invariance. This effect can be detected statistically by searching for the characteristic four point correlations. If the $\esttau$ estimator derived in the previous section were applied to the lensed CMB maps, it would also pick up significant spurious signal produced by lensing.  

We now quantitatively calculate the lensing bias to the $\esttau$ estimator. Lensing simply deflects the path of CMB photons from the last scattering surface resulting in a remapping of the CMB temperature/polarization pattern on the sky. The deflection angle $\bfd(\bn)$ is related to $\len(\bn)$, the lensing gravitational potential as
\begin{eqnarray}
d(\bn) = \nabla\len(\bn)\,. \label{eqn:dphi}
\end{eqnarray}
The lensing potential $\len(\bn)$ is given by
\begin{eqnarray}
\len(\bn)&=&- 2 \int_0^{\rad_0} d\rad
\frac{\da(\rad_0-\rad)}{\da(\rad)\da(\rad_0)} \Phi (\rad,\bn )
\, , \label{eqn:lenspotential}
\end{eqnarray}
where $d_A$ is the co-moving distance along the line of sight; $r_0$ is the comoving distance to the surface of last scattering, and $\Phi$ is gravitational potential~\cite{LewisChallinor2006}.

Similar to the effect of screening from patchy reionization, a lensing potential mode with wavevector $\bfl$ mixes the two polarization modes of wavevectors $\bfl_1$ and $\bfl_2=\bfl -\bfl_1$. Taking the ensemble average of the CMB fields for the fixed $\phi(\bn)$ field, similar to Eq.~(\ref{BasicDifference}), one gets
\begin{equation}
\langle \tilde X(\bfl_1)\tilde X'(\bfl_2) \rangle_{\rm CMB} = f^{\phi}_{XX'}(\bfl_1,\bfl_2) \len(\bfl)\,.
\label{BasicDifferenceLens}
\end{equation}
The form of filters $f^{len}_{XX'}(\bfl)$ for different combinations
of CMB field $X$ and $X'$ are given in Table~\ref{taufilter}. The
major difference between the filters of lensing potential estimator
$\hat\phi(\bn)$ and those for the $\hat\tau(\bn)$ estimator is the
additional factors of $\sim l^2$ that appear in those for lensing
owing to the differential in Eq.~(\ref{eqn:dphi}). This differential
nature of lensing significantly suppress the level of lensing
estimator noise since $l \gg 1$ and $N^{lens}_{XX'}(l)$ is
approximately proportional to $l^{-2}$ (see
Eq.~(\ref{eq:noise})). 
%The variance of the $\esttau$ estimator scales as $N(l)\propto 1/l^2$,  which can be understood by noting that (1) the noise is inversely proportional to the filter $N(l)\propto 1/f^2(l_1,l_2)$ (Table~\ref{taufilter}), and (2) for small $l$, $l_1 \approx l_2$ and hence $\sin(\varphi_{l_1 l_2}) \approx \varphi_{l_1 l_2} \propto l/l_1$. For lensing, the filters $f_{lens}$ contains additional factors of $l\cdot l_1$ because of the differential nature of lensing.  These additional factors result in a lensing estimator noise being much smaller than the $\tau(\bn)$ estimator noise.  Thus the estimator noise for $\tau(\bn)$ reconstruction is much larger than that for lensing reconstruction. 

%The remapping of observed CMB photons due to gravitational lensing induce a characteristic mode coupling.  This mode-coupling can mimic the patchy reionization signal, biasing $\tau(\bn)$ reconstruction.
 Let us consider a CMB sky that has been modified by both inhomogeneous reionization and lensing.  Suppose we want to reconstruct $\tau(\bn)$ assuming that all of the non-Gaussianity is from patchy reionization, which is equivalent to applying the $\esttau$ estimator with filters designed to optimally reconstruct $\tau(\bn)$. In this case, the estimator measures
\begin{eqnarray}
\label{eq:bias}
&&\langle \hat \tau_{XX'}({\bfl}) \rangle_{\rm CMB} = \tau({\bfl})
  \nn\\ && + \underbrace{N_{XX'}(l) \intl{1} f^{\phi}_{XX'}F^{\tau}_{XX'}(\bfl_1,\bfl_2)
  \len(\bfl)}\,, \\
&&\hspace{3.5cm} \text{bias}\nonumber
\end{eqnarray}
where $N_{XX'}$ is given by Eq.~(\ref{eq:noise}). The first term on
the right hand side is the desired signal, and the second term is a bias that owes to lensing.  Note that $f^{\phi}_{XX'}$ is the lensing filter (see Table~\ref{taufilter}) and $F^{\tau}_{XX'}(\bfl_1,\bfl_2)$ is given by Eq.~(\ref{filter1}) and (\ref{filter2}).

We simulate a patchy reionization induced $\tau(\bn)$ field (model B
in Table~\ref{taumodel}) and modulate the CMB fields by the
$\tau(\bn)$ field accordingly. We compare the reconstructed
$\tau(\bn)$ with the input field in Fig.~\ref{fig2} (see
Appendix~\ref{sec:simulation} for details of the simulations).  The
$\esttau$ estimator is unbiased if primordial CMB fields were unlensed
and only affected by patchy reionization.  However, in the presence of
lensing, the reconstructed $\tau(\bn)$ deviates significantly from the
fiducial signal. The lensing induced non-Gaussianity is rougly an
  order of magnitude larger than the patchy reionization induced
  non-Gaussianity.
%The $\esttau$ estimator picks up lensing non-Gaussianity that is much larger than the patchy reionization signal in our model.

The lensing induced non-Gaussianity could be reduced by applying lensing estimator $\hat\phi(\bn)$ to reconstruct the lensing potential, and then ``remap'' the observed CMB photons given the reconstructed $\phi(\bn)$ and Eq.~(\ref{eqn:dphi}). This process of subtracting the lensing effect from CMB is referred to as ``delensing'' (see Ref~\cite{CMBPolLensing} for a review). To investigate the lensing bias after applying this delensing procedure, we assume the residual lensing potential power spectrum is only $2\%$ of the fiducial value. The delensing fraction taken here is smaller than the predicted delensing fraction for future CMB experiment, using lensing maps either externally reconstructed from large scale structure/CMB temperature or from CMB polarization itself~\cite{Smith2010}. We find that even after delensing on the CMB map, the reconstructed $\tau({\bn})$ field is still significantly contaminated by the residual lensing signal, as shown in Fig.~\ref{fig2}. 
%more discussion on delensing

In Fig.~\ref{reconstructedPower}, we show the reconstruction of
optical depth fluctuation power spectrum $\hat C_{l}^{\tau \tau}$, and
compare with the input power spectrum $C_{l}^{\tau \tau}$. Again we
choose Model B for the reionization simulations, which has the highest
level of $\tau(\bn)$ fluctuations. We find that the lensing induced
spurious signal dominates over the fiducial signal by
$\sim\mathcal{O}(10-100)$, especially for $l\gtrsim200$. The
theoretical prediction for the spurious patchy reionization signal
from lensing which is calculated by Eq.~(\ref{eq:bias}), matches well with
$\hat C_{l}^{\tau \tau}$ from the simulation. Finally we show that
even after applying the delensing procedure with lensing quadratic
estimator~\cite{HuOkamoto}, the reconstructed $\hat C_{l}^{\tau \tau}$
is still biased by a factor of $\sim 10$. As we show in
Fig.~\ref{reconstructedPower}, the lensing induced
$C_{l}^{\tau\tau}$ has two bumps one peak at large scale $l\sim 50$
and the other peaks at small scale $l\sim 1000$. It is caused by the
lensing bias given by Eq.~(\ref{eq:bias}) is negative at low $l$ and
positive at high $l$ with a transition at $l\approx 200$. This sign change is 
because the lensing bias involves the product of the lensing and tau
filters [see Eq.~(\ref{eq:bias})]. The product contains a mode
coupling term $\mathbf{l} \cdot \mathbf{l}_1$ which is caused by the
derivative nature of lensing and gives the negative contribution at
low $l$. Physically, the lensing of CMB does not generate new power in
the CMB fluctuations, it only move power from large scale to small
scales~\cite{LewisChallinor2006}. We note that in principle lensing
reconstruction is also biased by the patchy reionization induced
non-Gaussianity, however since lensing signal is much larger than the
patchy reionization signal, we don't expect a significant
comtanimation from patchy reionization to lensing esitmation.
%the bias is at the sub-$\%$ level. Can we double check this by having something like fig 3 but now estimating bias vi tau on lening?}

%{\bf [Explain why this happens...I would have thought it would always be positive since it's generating non-gaussian noise....unless somehow it is canceling the nongaussianity in tau, which seems strange?  Also, can you change the linestyle when it goes negative.  Need to say you are looking at the left panel here. } Answer: The bias from lensing is given by Eq.(14), which shows that the bias term involves the product between the lensing and tau filter. The only term which can give negative contribution to bias should come from: $\mathbf{L} \cdot \mathbf{l}_1$. 
% Answer: We do include instrumetal effects (noise and beam). ``Delensing using the quadratic estimator''  just means that we are using the quadratic estimator for lensing to reconstruct the lensing signal $\hat \phi(n)$ (see Smith et al. arXiv:0811.3916 for details). The CMB is then de-lensed using $\hat \phi(n)$.}

\begin{figure*}[t]
\includegraphics[width=88mm,angle=0]{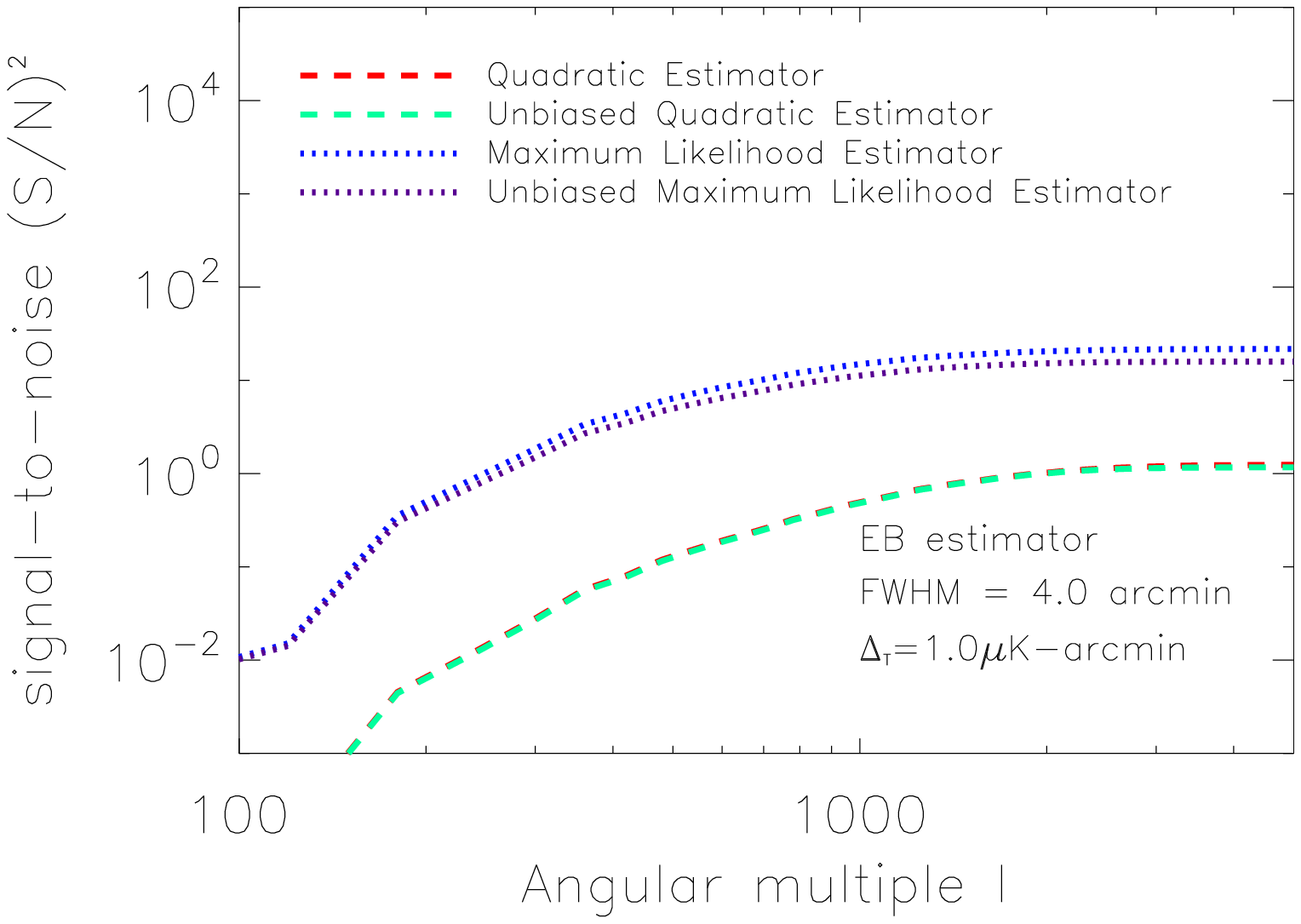}
\includegraphics[width=88mm,angle=0]{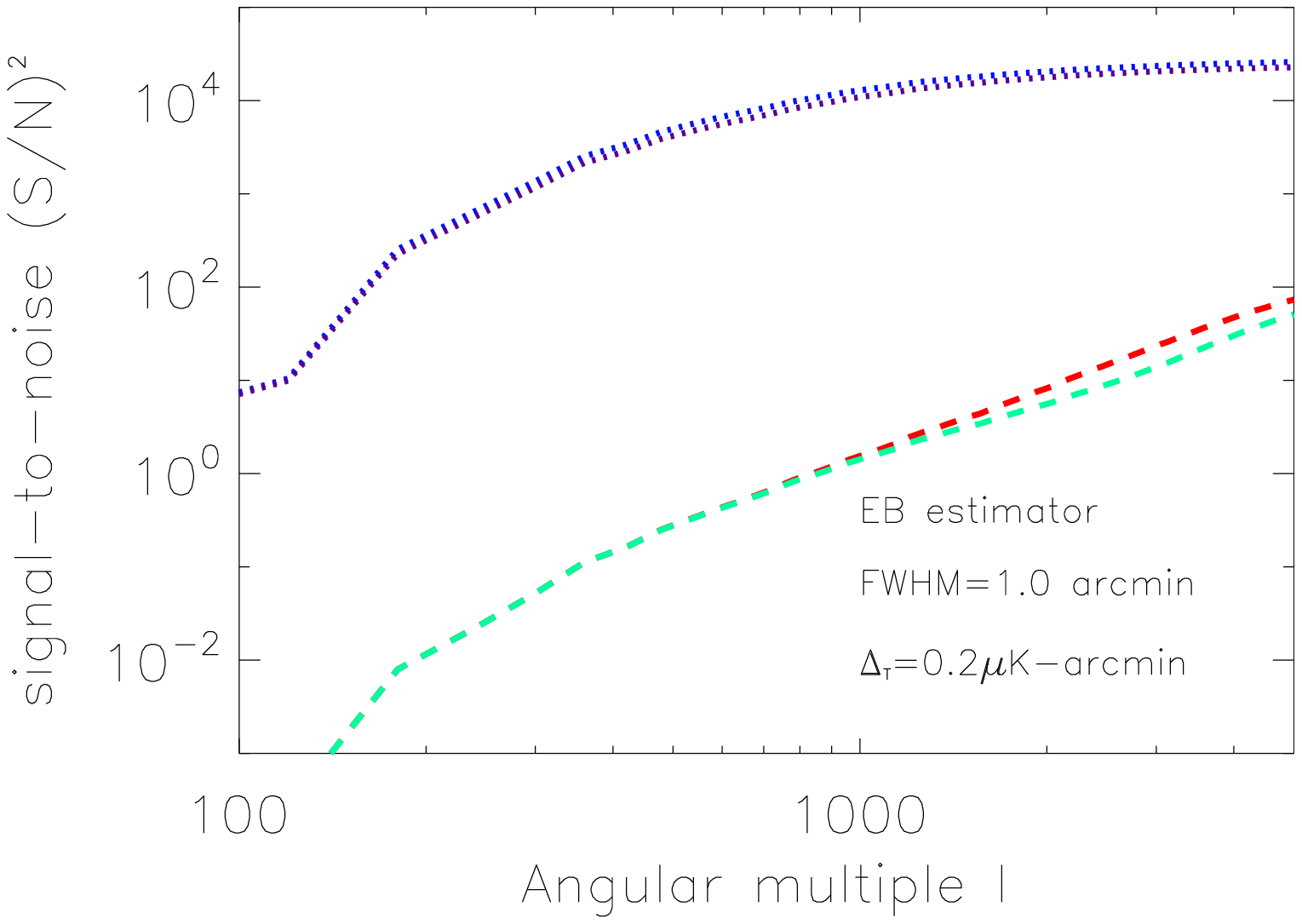}
\caption{The cumulative $(S/N)^2$ for the $EB$ estimator as a function of maximum multipole $l$. The left panel is for a CMBPol-like experiment with a beam of $\Theta_{\rm FWHM} =4'$ and the noise sensitivity $\Delta_p=1\mu K$-arcmin. The right panel is a more sensitive experiment with $\Theta_{\rm FWHM} =1'$ and $\Delta_p=0.2\mu K$-arcmin.  In both panels, we use the optical depth power spectrum $C_{l}^{\tau\tau}$ of model B in Fig.1. The green/red dotted curves (lower two curves) which almost overlap are the $(S/N)^2$ of the biased/unbiased quadratic estimators, whose noise level is given by Eq.~(\ref{eq:noise}) and~(\ref{eqn:unbiasedNoise}) respectively. The magenta/blue dashed curves (the top two curves that nearly overlap ), are the $(S/N)^2$ of the biased/unbiased maximum-likelihood estimators. The maximum-likelihood estimator is calculated from the minimum variance quadratic estimator except with the lensed CMB power spectrum replaced by the primary CMB power-spectrum (without the presence of lensing or patchy reionization effects) as suggested in~\cite{HirataSeljak2003}.}
\label{fig:unbiasednoise} 
%\end{center}
\end{figure*}

% The right panel in Fig.~(\ref{reconstructedPower}) demonstrates how lensing affects the $(S/N)^2$ as a function of multipole.  If the $\esttau$ estimator is blindly applied to the lensed CMB field, one is guaranteed to have a detection with $(S/N)^2 \sim10^5$ for a reference future experiment with $\Theta_{\rm fwhm} =1'$ and $\Delta_p=0.2\mu K$-arcmin. However, the detection has nothing to do with patchy reionization which only gives $(S/N)^2 \sim10^2$ . The detection is simply due to lensing leakage, and the typically spurious $(S/N)^2$ is $\sim \mathcal{O}(10)$ to $\sim \mathcal{O}(100)$ larger than the patchy reionization signal. 

%\section{Reconstructing Patchy Reionization}

\section{Reconstructing  Patchy reionization}
\label{sec:unbiase}

\subsection{Unbiased Estimator}
This section constructs an unbiased estimator for $\tau(\bn)$. As with the quadratic estimator discussed in the previous section, among all the six estimators, the $EB$ estimator has the highest $S/N$ ratio, thus we focus on $EB$ estimator in this section. For each multipole $l$ we can define a $2$-by-$2$ Fisher matrix $F^{\alpha\beta}_l$,
\ba
F^{\alpha \beta}_l &=& \intl{1}f^{(\alpha)}_{EB}(l_1,l_2)
({\bf C}^{-1})^{EE}_{l_1}f^{(\beta)}_{EB}(l_1,l_2)
({\bf C}^{-1})^{BB}_{l_2} \nonumber 
\ea
where $\alpha$ and $\beta$ run over  $\tau$ and $\phi$. 
The element $(F^{(-1)})^{\alpha \beta}_l$ of the inverse of Fisher matrix gives the variance of $C_{l}^{\alpha\beta}$. Hence the variance of $C_{l}^{\tau\tau}$ is:
\ba
N(l) \equiv \left({\bf F}^{-1}\right)_{l}^{\tau\tau} &=& \frac{F_l^{\phi\phi}}{F_l^{\tau\tau}F_l^{\phi\phi}-(F_l^{\tau\phi})^2} .
\label{eqn:unbiasedNoise}
\ea
This is the Gaussian noise term of the unbiased $\esttau$ estimator, which we will use to calculate $(S/N)^2$ in Fig.~\ref{fig:unbiasednoise}.

Starting from biased estimator $\esttau$ and $\estphi$~\cite{HuOkamoto}, we have
\ba
\langle \hat \tau({\bfl}) \rangle_{\rm CMB} &=& \tau({\bfl}) +  \frac{F_l^{\tau\phi}} {F_l^{\tau\tau}}\phi({\bfl})~, \nn \\
\langle \hat \phi({\bfl}) \rangle_{\rm CMB} &=& \phi({\bfl}) +  \frac{F_l^{\phi\tau}} {F_l^{\phi\phi}}\tau({\bfl}) \,.
\ea
One can then solve above equations for $\tau(\bfl)$ and $\phi(\bfl)$
\ba
\tau({\bfl}) &=& \frac{F_l^{\phi\phi}F_l^{\tau\tau}\langle \hat \tau({\bfl}) \rangle_{\rm CMB} - F_l^{\tau\phi}F_l^{\phi\phi}\langle \hat \phi({\bfl}) \rangle_{\rm CMB}}{F_l^{\tau\tau}F_l^{\phi\phi}-(F_l^{\tau\phi})^2}, \nn \\
\phi({\bfl}) &=& \frac{F_l^{\phi\phi}F_l^{\tau\tau}\langle \hat \phi({\bfl}) \rangle_{\rm CMB} - F_l^{\tau\phi}F_l^{\tau\tau}\langle \hat \tau({\bfl}) \rangle_{\rm CMB}}{F_l^{\tau\tau}F_l^{\phi\phi}-(F_l^{\tau\phi})^2}.
\ea
This estimator although unbiased is not a minimum variance estimator. In next subsection we compare the variance (Gaussian noise) of the minimum variance quadratic estimator with the variance of the unbiased estimator and show that there is only a marginal increase in the variance of the unbiased estimator in comparison to the variance of the minimum-variance estimator. 

{\it Maximum Likelihood Estimator:} Given that the $B$-mode polarization is well mapped, \citet{HirataSeljak2003} found that for lensing reconstruction the maximum-likelihood estimator (which reduces the estimator noise from lensing) allows significantly better $(S/N)^2$ than the quadratic estimator. 
%The maximum likelihood method formally allows one to perfectly reconstruct $\tau$ from $EB$ with an ideal measurement {\bf [because....]}

Following Ref.~\cite{HirataSeljak2003}, the lensing  maximum-likelihood estimator can be generalized to construct a unbiased maximum-likelihood estimator for $\tau(\bn)$ in the presence of lensing. The variance of the maximum-likelihood estimator for $\tau(\bn)$ is the same as that for the quadratic estimator $\esttau$ with one exception--- for maximum-likelihood estimator the denominator of Eq.~(\ref{filter1}) and Eq.~(\ref{filter2}) contains the unlensed CMB power spectrum, whereas the quadratic estimator noise contains the lensed CMB power spectrum~\cite{HirataSeljak2003}. The estimator noise of $\tau(\bn)$ reconstruction would no longer be saturated because of the lensed CMB power spectrum. Conceptually, the lensing or patchy reionization induced B-modes can be iteratively cleaned from the map, therefore we are able to reduce the post-cleaning B-mode power spectrum and thus reducing the noise in the $\esttau$ estimator. Our fundamental ability to clean the map is bounded by the sum of the unlensed CMB B-modes and the instrumental noise.  

\begin{figure*}[t]
\includegraphics[width=60mm,angle=-90]{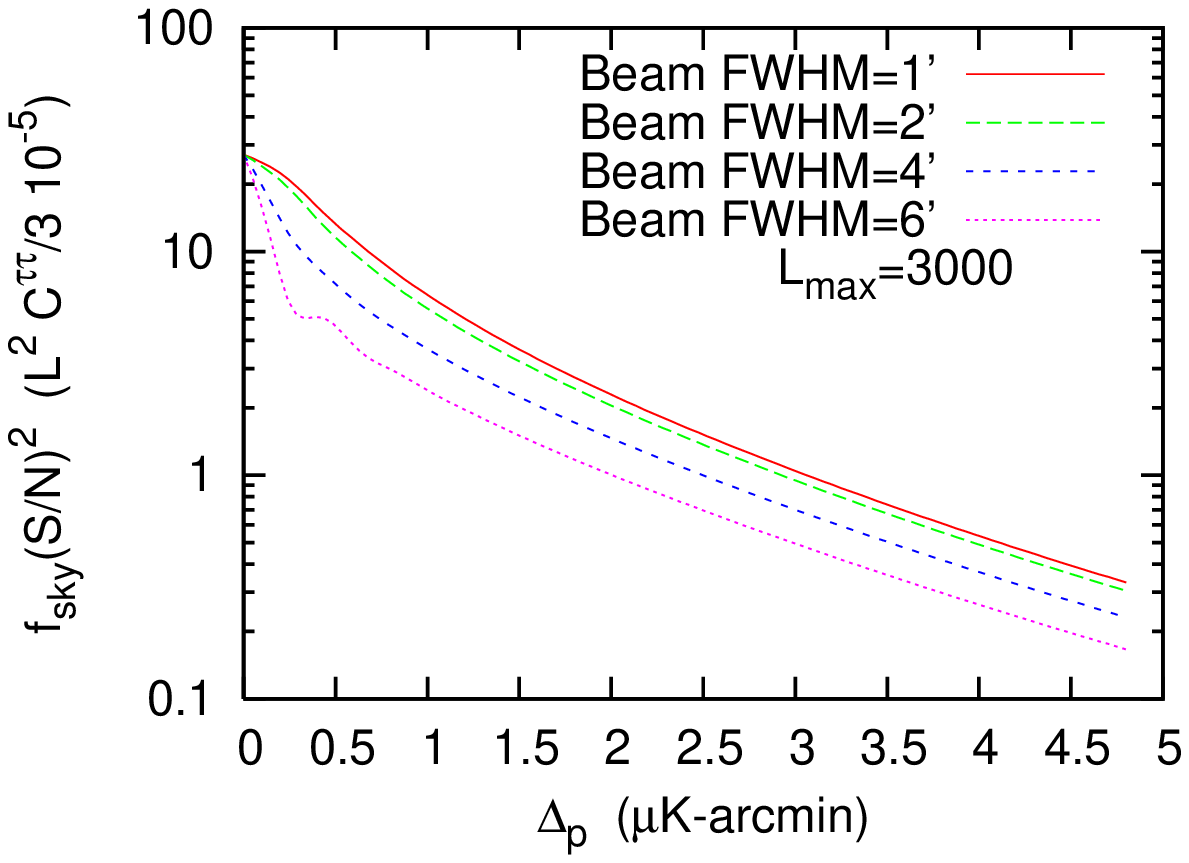}
\includegraphics[width=60mm,angle=-90]{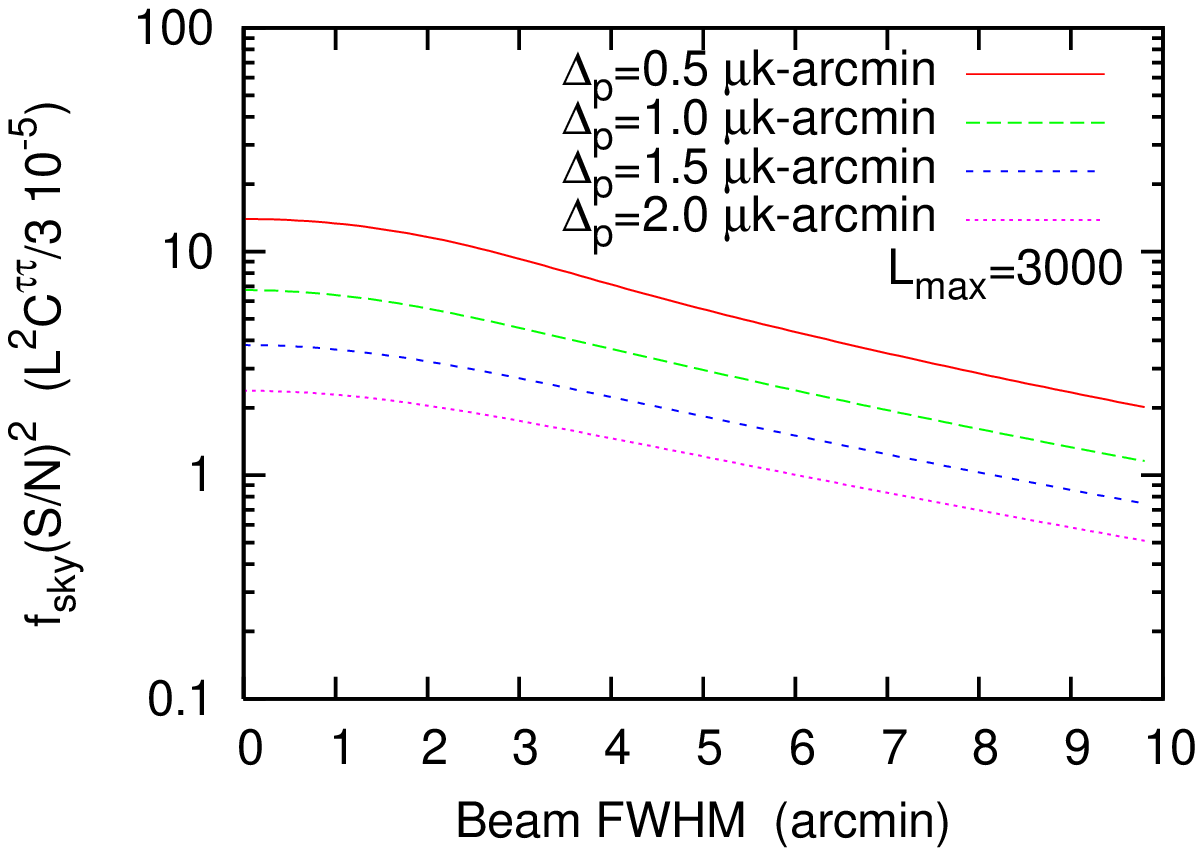}
\caption{Dependence of the total $(S/N)^2$ on instrumental sensitivity, $\Delta_p$, (left panel) and beam size, $\Theta_{\rm FWHM}$ (right panel).  We plot $EB$ unbiased quadratic estimator as an example and calculate cumulative $(S/N)^2$ up to $l_{max}=3000$. {\it Left panel}, we fix beam size with 1, 2, 4, and 6 arcmin respectively to calculate the $(S/N)^2$ dependence on instrumental sensitivity. {\it Right panel}, we fix $\Delta_p$ with 0.5, 1.0, 1.5, and 2.0$\mu$K-arcmin to show the beam size dependence of $(S/N)^2$. The approximate fitting functions are given in Eq.~(\ref{dependence}).}
\label{fig:instrument} 
%\end{center}
\end{figure*}

\subsection{Forecasting the Detectability of Patchy Reionization}

The signal-to-noise for the detection of patchy reionization signal can be written as
\be
\Big(\frac{S}{N}\Big)^2= \Bigg[ \frac{f_{sky}}{2} \sum_l (2l+1) \Bigg(\frac{C^{\tau \tau}_{l}}{N(l)}\Bigg)^2\Bigg],
\ee
where $f_{sky}$ is the sky fraction; $C^{\tau \tau}_{l}$ is the fiducial patchy reionization power spectrum, and $N(l)$ is the leading order Gaussian noise of an estimator, given by Eq.~(\ref{eqn:unbiasedNoise}) for the unbiased quadratic estimator and given by Eq.~(\ref{eq:noise}) for the biased minimum variance quadratic estimator.

In Fig.~\ref{fig:unbiasednoise}, dashed-lines (the lower two curves
which almost overlap) compare the $(S/N)^2$ of the biased and unbiased
quadratic estimators. The left panel is for the CMBPol like experiment
with  with noise $\Delta_T=1 \mu \text{k-arcmin}$ and beam FWHM
$\Theta_{\rm fwhm}=4$ arcmin. The right panel is for the reference experiment with noise $\Delta_T=0.2 \mu \text{k-arcmin}$ and beam FWHM $\Theta_{\rm fwhm}=1$ arcmin. As is clear from figure the $(S/N)^2$ of unbiased estimator is only slightly lower than the $(S/N)^2$ of biased estimator for both CMBPol like experiment and the reference experiment. In another word, the variance of the unbiased estimator is only marginally more (percent-level) than the variance of the minimum-variance quadratic estimator. The reason for this is easy to understand--- the contribution to the variance from the spurious $\tau(\bn)$ signal produced by lensing is much smaller than the intrinsic $\esttau$ estimator noise.

In Fig.~\ref{fig:unbiasednoise}, dotted-lines (the upper two lines which almost overlap) compare the $(S/N)^2$ of the biased/unbiased maximum-likelihood $\hat \tau(\bfl)$ estimators. For CMBPol-like experiment, the maximum-likelihood estimator can get $(S/N)^2$ about a factor of $10$ higher than the quadratic estimator. 
 
%However, we note that the noise of the maximum-likelihood estimator is optimistic.  We have not considered other secondary anisotropies like polarization generated from the quadrupole scattering in ionized bubbles, which would potentially affect the estimated noise of maximum-likelihood estimator. {\bf [give estimate for this.  It's just $C_\tau x10^{-5}$, where $10^{-5}$ is power in the incident quadrupole on each patchy.  I suspect that it may not be optimistic because this is so small.  Read B-modes seem like the biggest hindrance.  ]}

In Fig.~\ref{fig:instrument}, we show the total $(S/N)^2$ from the unbiased quadratic estimator as a function of instrumental beam size and detector sensitivity respectively. We find that the $(S/N)$ for a constant $l^2 C^{\tau \tau}_{l}/(2\pi)$ and for experiment with $\Delta_P > 1 \mu \text{k-arcmin}$ can be approximated as 
\begin{eqnarray}
\Bigg(\frac{S}{N}\Bigg)^2 &\approx& 5 f_{sky} \, \left( \frac{l^2 C^{\tau \tau}_{l}/2\pi}{5\times 10^{-6}} \right) \, \exp\left(\frac{-0.56\Delta_p}{1 \mu K-\text{arcmin}} \right)\, \nonumber \\ & &
\times \exp\left(\frac{-0.09\Theta_{\rm FWHM}}{1'} \right)\,.
\label{dependence}
\end{eqnarray}
The $(S/N)^2$ is more sensitive to the instrumental sensitivity rather than the beam size. For a reference pathcy reionization signal $l^2 C^{\tau \tau}_{l}/2\pi = 5\times 10^{-6}$, for a CMBPol-like or COrE-like~\cite{COrE} experiment we expect $(S/N)^2 \sim\mathcal{O}(1)$. For the future ground-based experiments such as the POLAR Array with $\Delta_p = 1.41 \mu$K-arcmin, $\Theta_{\rm FWHM} = 1'$, and sky coverage 100 deg$^2$, we expect $(S/N)^2 \sim\mathcal{O}(0.01)$.

%The imprint of reionization is most likely to be the most detectable imprint of reionization on the CMB.  The information about reionization in the kSZ is very similar to in $C_{\tau \tau}$.
%The kinetic Sunyaev Zeldovich anisotropies are given by
%\begin{equation}
%\delta T_{\rm kSZ} = \int d\eta \frac{d\tau(\bn)}{d\eta} \, \bn \cdot {\bf v} 
%\end{equation}
%In fact, its power spectrum contains very similar information to $C_{\tau \tau)$.  The power spectrum of the kSZ from reionization can be written as $C_{\rm kSZ) \approx \sigma_v(z_*)^2 \, C_{\tau \tau)$, where $\sigma_v(z_*)$ is the RMS line-of-sight velocity dispersion at the effective redshift of reionization, $z_*$, in units of $c$.

%It is also possible to reconstruct $\tau$ from the kSZ.  In particular, at large scales there is a lower order anisotropy called the Doppler anisotropy that comes from a homogeneous electron field and an inhomogeneous velocity field.  Since only line-of-sight velocity modes contribute, this anisotropy is strongly suppressed at $L > 100$, but is of order a tenth of the primary anisotropies at $L \sim 10-100$.  {\bf check}.  

\section{Summary and Discussion}
\label{sec:nume}

Reionization marks the epoch in which the vast majority of the hydrogen in the Universe was ionized since cosmological recombination.  When and how reionization occurred is at present poorly constrained.  In addition to pinning down the epoch of this cosmic phase transition, constraints on the reionization history provides us information about the formation of early galaxies. Inhomogeneous reionization would have generated fluctuations in the Thomson scattering optical depth $\tau(\bn)$ among different lines of sights at the level $\sim 10^{-3}$.  These modulations would modify the primordial CMB temperature and polarization anisotropies by inducing a directionally dependent screening.  Such screening couples different modes of CMB, converting $E$-modes to $B$-modes, and introduces non-Gaussian signals.  

% In our previous paper, we have shown that in general, any distortions on CMB field could be in principle optimistically reconstructed by applying a specifically designed quadratic estimator to the observed CMB fields. If we consider patchy reionization is a modulation field distort the primordial CMB field, then the line of sight dependent optical depth could be in principle be reconstructed. The question is how much the signal to noise is, and in reality how other effects affect such reconstruction.

 In this paper, we used a technique that exploits the non-Gaussianities in the CMB sourced by reionization to study this process, as first proposed in \citet{DvorkinSmith}. We have introduced the the minimum variance quadratic estimator in an intuitive flat sky limit and compared it with the estimator for lensing potential reconstruction~\cite{HuOkamoto}.  Lensing induced non-Gaussian features would produce a spurious $\tau(\bn)$ signal that is at least an order of magnitude higher than our semi-analytical models predict from patchy reionization.  We showed that ignoring the lensing contamination would significantly bias the reconstruction of optical depth fluctuation field $\tau(\bn)$.  Even after applying a delensing procedure that used the minimum variance quadratic estimator for the lensing potential $\phi(\bn)$, the residual lensing bias on the $\hat\tau(\bn)$ estimator was still comparable with the fiducial value.  As a solution, we constructed an unbiased estimator to simultaneously reconstruct $\tau(\bn)$ and the lensing potential $\phi(\bn)$ such that the estimate of $\tau(\bn)$ is not biased by lensing. We found that the $S/N$ of the unbiased estimator is only degraded at the percent level compared to the original biased $\tau(\bn)$ estimator.  

We studied the detectability of patchy reionization by considering
more detailed $\tau(\bn)$ fields using semi-numerical reionization
models, which unfortunately yield an order-of-magnitude smaller signal
than previously considered~\citep{DvorkinSmith}.  As a result, we found that with
the unbiased estimator, a CMBPol-like experiment could achieve a
marginal detection of patchy reionization with $(S/N)^2 \sim
1-10$. We characterized the estimator noise for various instrumental properties. We find that the $S/N$ is only weakly sensitive to the FWHM of detector beam with a factor of $\sim$2 degradation of $(S/N)^2$ by increase FWHM from $1'$ to $6'$. While the $(S/N)^2$ decreases by a factor of $\sim$2 by increase instrumental noise from 0.5 to 2 $\mu \text{k-arcmin}$. Similar scaling with instrumental characteristics have been quantified for lensing reconstruction in~\cite{HuOkamoto}. 

% as $(S/N)^2 \sim \text{exp}(\frac{-0.09 \Theta_{\rm  FWHM}}{1'})$. While the  $S/N$  scaling with instrumental noise goes as $(S/N)^2\sim \text{exp}(\frac{-0.56 \Delta_p}{1 \mu \text{k-arcmin}})$. {\bf [Can we write these expressions in words...This is too bulky in the text otherwise.  Also, should state why a weak scaling with beam.]}

Large scale CMB fields are also modulated by smaller scale $\tau(\bn)$
fluctuations due to patchy reionization. As we construct the $\esttau$ estimator in flat sky limit, we ignore the patchy reionization signal from large scale $E/B$-mode which is generated via scattering of the local CMB temperature quadrupole by ionized bubbles. The $S/N$ will be increased by a factor of $\sim$2 by considering such signal on large scales~\cite{DvorkinSmith}.

%We generalized the ideas from lensing to construct a maximum likelihood-based estimator for $\tau(\bn)$ and compared with the quadratic estimator $\hat\tau(\bn)$. The main distinction between this approach and the quadratic estimator is that the likelihood method replaces the observed CMB power spectrum in the weights of the quadratic estimator with primordial undistorted CMB power spectrum. The replacement can be understood as an iterative process that applies the quadratic estimator repeatedly.  The reconstruction of $\tau(\bn)$ is limited only by the instrument noise level.  For analytic estimate for the reconstruction uncertainty in maximum likelihood estimator we calculate the Fisher limit to provide the lower bound of the lens reconstruction error (see~\cite{HirataSeljak2003} for details).

Although the predicted $S/N$ for a patchy reionization detection is only marginal for a CMBPol-like experiment, one can cross-correlate with other cosmological data sets that are sensitive to the properties of patchy reionization.  The same population of ionized bubbles would not only induce line of sight dependent optical depth of CMB, but also correlate with the distribution of galaxies or the redshifted $21$cm signal.  

At large scales, it is expected that the distribution of galaxies
correlates well with the neutral gas distribution~\cite{zahn2007}.
One can estimate the $(S/N)_{\tau}$ for a patchy reionization detection as

\begin{equation}
(S/N)_{\tau}^2 = l_{\rm char}^2 (C^{\tau\tau}_{\rm char})^2/(C^{\hat \tau \hat \tau}_{l})^2
\end{equation}
where $l_{\rm char}$ is the characteristic multipole that contributes to the $S/N$ ($l_{\rm char} \sim 10^3$), and $C^{\hat \tau \hat \tau}_{l}$ is the $\esttau$ estimator variance.  The scaling factor $l_{\rm char}^2$ is an estimate for the number of modes that are contributing to the signal (the result really does not rely on the fraction of the sky $\tau(\bn)$ is estimated).

An estimate for signal-to-noise that can be obtained in cross correlation $(S/N)_{\tau g}$ is 

\begin{eqnarray}
(S/N)_{\tau g}^2   &\approx& l_{\rm char}^2 \, f_{\rm sky} \,f_{\rm reion} \, (C^{\tau g}_{\rm char})^2/(C^{\hat \tau \hat \tau}_{l} C_{gg})\, \nonumber \\
                 &=& l_{\rm char}^2 \, f_{\rm sky} \, f_{\rm reion} \, r^2\, C^{\tau\tau}_{\rm char}/C^{\hat \tau \hat \tau}_{l} \, \nonumber \\
                 &=& l_{\rm char} \, f_{\rm sky} \, f_{\rm reion} \, r^2\, (S/N)_\tau,\end{eqnarray}
where $f_{\rm sky}$ and $f_{\rm reion}$ are the fraction of the sky and reionization over which surveys overlap, $r$ is the cross correlation coefficient of galaxies and the $\tau(\bn)$ field over the same projected volume as the galaxy survey ($r \sim 1$ on large scale).

%We have ignored terms like cross correlation between estmiator variance and galaxy $C_{\rm var-g}$ in the denominator, which should be small since the estimator variance is coming from things not correlated with reionization-redshift galaxies. 
Noting that $f_{\rm sky} \lesssim 10^{-4}$ is the current size for $z\sim7$ galaxy surveys, it would take a very ambitious survey to enhance the $\tau$ signal in cross correlation compared to in the auto-power.  Correlating with the diffuse background light from early galaxies -- the cosmic infrared background -- is a related and intriguing possibility since then $f_{\rm sky} f_{\rm reion} \sim 1$ (although, lower redshift emission may be a significant noise source in this case) and may deserve further study.  

The final possibility that we discuss is cross correlating with a survey of redshifted $21$cm emission from intergalactic neutral hydrogen.  Such surveys do span a significant fraction of the sky and the first generation of such endeavors will be in a noise-dominated regime in which they could benefit from cross-correlation~\cite{Lidz2011} (Note that cross correlating with $\tau$ would be of little interest if there existed high $S/N$ $21$cm maps).  However, redshifted  $21$cm analyses remove the modes with small line-of-sight projected wavevectors in the act of foreground cleaning, which unfortunately are the modes that contribute to the $\tau$ signal~\cite{McQuinn2006}. Thus, there would be little signal in this cross correlation.

%{\bf [It would be nice to discuss qualitatively why the S/N is much smaller than for lensing: $\delta \tau \gg \delta \phi$, and it's these factors of $l$ from lensing.]} 

%The method discussed here can be applied to study ``other'' processes which distort CMB to generate 4-point functions. For example the kinetic Sunyaev Zeldovich (kSZ) signal will also distort CMB generating non-Gaussianities. Quadratic estimator formalism can be applied to constrain/detect kSZ power-spectrum. The power spectrum of kSZ anisotropies from during reionization $C^{\rm kSZ}_{l}$ contains very similar information to $C_{l}^{\tau \tau}$.  Because most of the power in the velocity field is at large scales, $C^{\rm kSZ}_{l}\approx \sigma_v(z_*)^2 \, C_{l}^{\tau \tau}$ at $l \gtrsim 100$, where $\sigma_v(z_*)$ is the RMS line-of-sight velocity dispersion at the effective redshift of reionization, $z_*$, in units of $c$. (Matt, I think you wrote this paragraph before would you like to say more about kSZ reconstruction?)  

% It has been shown that this spatial dependent optical depth field $\tau(\bn)$ can in
% principle be reconstructed by quadratic estimators analogous to lensing reconstruction~\cite{DvorkinSmith}. 

% In this paper, we presented explicit simple formulae for estimators of the spatially varying optical depth $\esttau$ in the flat sky limit. We have shown that gravitational lensing highly bias the patchy reionization signal.

\acknowledgements{We thank C. Dvorkin and K. M. Smith for helpful discussions.  APSY gratefully acknowledges support from IBM Einstein fellowship and funding from NASA award number NNX08AG40G and NSF grant number AST-0807444.  MM is supported by the NASA Einstein fellowship. JY is supported by the SNF Ambizione grant. MZ is supported by the National Science Foundation under PHY-0855425 and AST-0907969, and by the David and Lucile Packard Foundation and the John D. and Catherine T. MacArthur Foundation.}

\bibliography{myreference}

\appendix

\section{Reconstructing Inhomogeneous Reionization $\tau({\bn})$: Simulation Pipeline}
\label{sec:simulation}

Our simulation pipeline of optical depth $\tau({\bn})$ field reconstruction follows the procedure in \citep{HuDeDeoVale}, and we modified the code developed for lensing reconstruction in \citep{YooZaldarriaga2008,YooZaldarriagaHernquist2010}. First, we generate primordial CMB polarization $Q^{pri}(\bn)$ and $U^{pri}(\bn)$ maps as Gaussian realizations of CMB power spectrum. We choose a standard fiducial model with a flat $\Lambda CDM$ cosmology, with parameters given by $\Omega_b=0.045, \Omega_c=0.23, H_0=70.5, n_s=0.96, n_t=0.0,$ and $\tau=0.08$. We calculate the theoretical lensed and unlensed CMB power spectrum from publicly available code CAMB~\cite{CAMB}. The primordial CMB polarizations maps are then transformed according to Eq.~(\ref{modulation}) to include the effect of patchy reionization. The $\tau(\bn)$ field was generated from a reionization simulation described in Section~\ref{sec:simu}. 

To include the effect of lensing, we generate a realization of lensing deflection field $\bfd(\bn)$ and transform the CMB fields $\tilde Q(\bn)$ and $\tilde U(\bn)$ to $Q(\bn)$ and $U(\bn)$ according to
\begin{eqnarray}
 (Q\pm iU)(\bn) &=& (\tilde Q \pm i\tilde U)(\bn+\bfd(\bn)) \,. 
\end{eqnarray}
The deflection angle at each point $\bn$ is calculated by taking the gradient of the lensing potential. The lensing potential power spectrum is generated using CAMB which was run with nonlinear corrections using halofit~\cite{CAMB}.

Since we want to quantify the lensing contamination, we have several pipelines with different level of lensing signal being removed. We define a de-lensing factor $\alpha$, as $C^{delen, \phi \phi}$=$C^{theory, \phi \phi}/\alpha$, where $\alpha=1$ correspond to no de-lensing, $\alpha\rightarrow \infty$ corresponds to perfect delensing, we use $\alpha=50$ for Fig.~(\ref{fig2}). 

We then Fourier transform CMB polarization maps to get $E(\bfl)$ and $B(\bfl)$ maps. Finally we multiply CMB $E(\bfl)$ and $B(\bfl)$ maps by Gaussian beam in Fourier space and add instrumental noise.

More specifically, we closely follow Hu et al~\cite{HuDeDeoVale} to re-write the $\esttau$ estimator which is more efficient to evaluate computationally. We re-write the estimator in real space
\begin{equation}
\hat \tau^{EB}_{\bfl} =-N_{l}^{EB}\int {d^2\bn } e^{-i \bn \cdot \bfl} {\rm Re} \left\{
[  {\bf G}^{EB} (\bn) L^{B*} (\bn)]  \right\}\,.
\label{eqn:fourierestimator}
\end{equation}
The field ${\bf G}^{EB}$ is built from the observed $E(\bfl)$ field (including contributions from lensing and patchy reionization) as
\begin{equation}
{\bf G}^{EB}_{\bfl} =  \frac{C^{EE}_l}{ ( C_{l}^{EE}+N_{l}^{EE})} {E(\bfl)}e^{2i\varphi_{\bfl}} \,.
\end{equation}
and
$L^B$ is given by 
\begin{equation}
L^{B}_{\bfl} =  \frac{B(\bfl)}{ ( C_{l}^{BB}+N_{l}^{BB})}e^{2i\varphi_{\bfl}} \,.
\end{equation}

$N_{l}^{EB}$ is a normalization coefficient which ensure the unbiansdness of the estimator~\cite{HuDeDeoVale}. The results of our simulations are shown in Fig.~(\ref{fig2}). 

\section{Unbiased Minimum Variance Quadratic Estimator for Patchy Reionization}
\label{sec:simulation}
This section discusses the quadratic estimator for the patchy
reionization induced optical depth fluctuation field $\tau(\bn)$ in a more general context,
demonstrating that the estimator $\esttau$ used in the text is the minimum
variance estimator in the limit that the signal-to-noise ratio in
lensing estimator $\hat \phi (\bn)$ is much higher than in $\hat \tau (\bn)$.   
%\begin{equation}
%\langle X(\bfl-\bfl_i) X'(\bfl_i) \rangle_{\CMB} = f_{\tau}(\bfl - \bfl_i, \bfl_i) \phi_\bfl + f_{\phi}(\bfl - \bfl_i, \bfl_i) \tau_\bfl.
%\end{equation}
%{\bf I expect we don't need this and should just refer back}

In Fourier space, the general unbiased quadratic estimator for $\hat\phi_\bfl$ and $\hat\tau_\bfl$ is
\begin{eqnarray}
\hat\tau_\bfl = \sum_{\bfl_i} Q_{\bfl_i}  X(\bfl-\bfl_i) X'(\bfl_i) - \Tr(Q f^{\phi}_{XX'} )\,\phi_\bfl \nonumber \\
\hat\phi_\bfl = \sum_{\bfl_i} P_{\bfl_i}  X(\bfl-\bfl_i) X'(\bfl_i) - \Tr(P f^{\tau}_{XX'} )\,\tau_\bfl
\end{eqnarray}
where $Q_{\bfl_i}$ and $P_{\bfl_i}$ are some weighting functions, $f^{\tau}_{XX'}$ and $f^{\phi}_{XX'}$ is the same as in
Eq.~(\ref{BasicDifference}) and (\ref{BasicDifferenceLens}). The sum
does not include $\bfl_i = 0$ and we are using $\Tr(X)$ as shorthand
for $\sum_{\bfl_i} X_{\bfl_i}$.  Noting that $\langle X(\bfl-\bfl_i)X'(\bfl_i)\rangle=f^{\phi}_{XX'} (\bfl-\bfl_i,\bfl_i) \phi_\bfl+f^{\tau}_{XX'} (\bfl-\bfl_i,\bfl_i) \tau_\bfl$,
we can write the above equation as (if we substitute the unbiased estimator $\hat\phi_\bfl$ and $\hat\tau_\bfl$ on the R.H.S.)
\begin{equation}
\mathbf{A} \times
\left( \begin{array}{c}
\hat{\tau}_\bfl \\ \hat{\phi}_\bfl \end{array} \right) = \left( \begin{array}{c}
\sum_{\bfl_i} Q_{\bfl_i}  X(\bfl-\bfl_i) X'(\bfl_i) \\ \sum_{\bfl_i} P_{\bfl_i} X(\bfl-\bfl_i) X'(\bfl_i) \end{array} \right)
\label{eq:eb}
\end{equation}
where
\begin{equation}
\mathbf{A} \equiv \left( \begin{array}{cc}
1 & \Tr(Q f^{\phi}_{XX'} ) \\ \Tr(P f^{\tau}_{XX'} ) & 1 \end{array} \right). 
\end{equation}

Thus, the general unbiased quadratic estimator for $\tau_\bfl$ alone is 
\begin{equation}
\hat\tau_\bfl = \sum_{\bfl_i} \left[[\mathbf{A}^{-1}]_{11}\,Q_{\bfl_i} + [\mathbf{A}^{-1}]_{12} P_{\bfl_i} \right] \, T(\bfl-\bfl_i) T(\bfl_i) 
\end{equation}

We want to derive the weighting functions $Q_{\bfl_i}$ and $P_{\bfl_i}$ that give us the unbiased minimum variance estimator $\hat\tau_\bfl$.  The estimator variance is
\begin{eqnarray}
{\rm var}[\hat\tau_\bfl]  &=&  \langle |\hat\tau_\bfl|^2\rangle  \nn \\
&=&  2 \, \sum_{\bfl_i} X_{\bfl_i} C_{\bfl_i} X_{\bfl_i} C_{\bfl - \bfl_i}\,,  
\label{B6}
\end{eqnarray}
where
\begin{equation}
X _{\bfl_i} =  [\mathbf{A}^{-1}]_{11} Q_{\bfl_i} + [\mathbf{A}^{-1}]_{12} P_{\bfl_i}\,.  
\label{B7}
\end{equation}
%(Technically the angle $\phi_0$ of the $\bfl=0$ mode is undefined, howeverthis will not concern us since within the flat-sky approximation we will
%convert sums over $\bfl$ into integrals: $\sum_\bfl\rightarrow\int d^2\bfl/\pi$.  If an integral is divergent at ${\bfl}=0$, then it cannot
%be computed accurately within the flat-sky approximation.)  {\bf Matt: why is this here?}

To derive the minimum variance estimator, we want to minimize ${\rm var}[\hat\tau_\bfl] $ subject to the conditions that $\Tr[Pf^{\tau}_{XX'} ] = 1$ and $\Tr[Qf^{\phi}_{XX'} ] = 1$.  Rather than go through this exercise, let us note first that at relevant multipole $f^{\tau}_{XX'}  \ll f^{\phi}_{XX'} $ because of the factor of $\sim l^2$ that contributes to $f^{\phi}_{XX'} $.  Let us also note that the weighting function $P_{\bfl_i}$ that is optimal for simultaneously estimating $\phi_\bfl$ with $\tau_\bfl$ should be nearly identical to the minimum variance quadratic weighting for estimating just $\phi_\bfl$ because $\tau_\bfl$ is a weak contaminant of lensing.  Second note that $|[{\bf A}^{-1}]_{12}| \propto \Tr[Pf^{\tau}_{XX'} ] \ll 1$ (since $\Tr[Pf^{\phi}_{XX'}] = 1$), and thus $[{\bf A}^{-1}]_{12}$ has magnitude that is much less than that of $[{\bf A}^{-1}]_{11}$. Not only $[{\bf A}^{-1}]_{11} \gg [{\bf A}^{-1}]_{12}$, but note the scaling $Q_{\bfl}/P_{\bfl}\sim l^2 \gg 1$, therefore, we are justified in ignoring the second term in Eq.~(\ref{B7}) and one can show that the minimizing Eq.~(\ref{B6}) subject to the constraint $\Tr[Qf^{\tau}_{XX'} ] = 1$ yields
\ba
Q_{\bfl_i} = \Tr[C^{-1} f^{\tau}_{XX'}  C^{-1} f^{\tau}_{XX'} ]^{-1} C_{\bfl_i} ^{-1} f^{\tau}_{XX'} (\bfl_i, \bfl - \bfl_i) C_{\bfl - \bfl_i}^{-1},\nn\\
\ea
which yields the identical estimator to that used in the text. Furthermore, because the variance of $\hat \tau_\bfl$ is dominated by $Q_{\bfl_i}$, this explains why our unbiased estimator that accounts for $\phi_\bfl$ yields a result that is not much different than the biased minimum variance estimator.

\end{document}